\documentclass[twocolumn,prb,superscriptaddress]{revtex4-2}

\usepackage{amsmath,amsthm,amssymb,mathtools,braket,amscd}
\usepackage{graphicx,parskip}
\usepackage{tikz}
\usepackage[many]{tcolorbox} 
\usepackage{hyperref}
\hypersetup{
    colorlinks = true,
    citecolor = blue,
    linkcolor = blue,
    urlcolor = teal,
    filecolor = teal,    
}
\theoremstyle{definition}

\newcommand{\X}{\sigma^x}
\newcommand{\Y}{\sigma^y}
\newcommand{\Z}{\sigma^z}

\def\RB{\textcolor{black}}

\begin{document}
\title{Symmetry-protected topological corner modes in a periodically driven interacting spin lattice}

\author{Kelvin Koor}
\email{cqtkjkk@nus.edu.sg}
\affiliation{Centre for Quantum Technologies, National University of Singapore, 3 Science Drive 2, Singapore 117543}

\author{Raditya Weda Bomantara}
\email{raditya.bomantara@kfupm.edu.sa}
\affiliation{Centre for Quantum Engineered Systems, School of Physics, University of Sydney, Sydney, New South Wales 2006, Australia}
\affiliation{Department of Physics, King Fahd University of Petroleum and Minerals, 31261 Dhahran, Saudi Arabia}

\author{Leong Chuan Kwek}
\email{kwekleongchuan@nus.edu.sg}
\affiliation{Centre for Quantum Technologies, National University of Singapore, 3 Science Drive 2, Singapore 117543}
\affiliation{National Institute of Education, Nanyang Technological University, Singapore 637616}
\affiliation{MajuLab, CNRS-UNS-NUS-NTU International Joint Research Unit, UMI 3654, Singapore}
\affiliation{Quantum Science and Engineering Centre (QSec), Nanyang Technological University, Singapore}

\begin{abstract}
Periodic driving has a longstanding reputation for generating exotic phases of matter with no static counterparts. This work explores the interplay between periodic driving, interaction effects, and $\mathbb{Z}_2$ symmetry that leads to the emergence of Floquet symmetry protected second-order topological phases in a simple but insightful two-dimensional spin-1/2 lattice. Through a combination of analytical and numerical treatments, we verify the formation of corner-localized 0 and $\pi$ modes, i.e., $\mathbb{Z}_2$ symmetry broken operators that respectively commute and anticommute with the one-period time evolution operator, as well as establish the topological nature of these modes by demonstrating their presence over a wide range of parameter values and explicitly deriving their associated topological invariants under special conditions. Finally, we propose a means to detect the signature of such modes in experiments and discuss the effect of imperfections. %Since strongly-interacting systems are ubiquitous in nature, it is natural to consider the effects of periodic driving on such systems, where exotic phenomena could be probed. Higher-order topological phases, which exist only at the corners of the lattice, is one such example. In this work we demonstrate the existence of second-order $\mathbb{Z}_2$-symmetry protected topological phases by periodically driving and dimerizing an interacting two-dimensional spin-1/2 system. The key idea in the construction of such a system is to start with vertically stacked copies of the well-known 1D Ising spin chain, and then implementing suitable vertical interactions, dimerization, and periodic driving step-by-step to obtain the final model. Using various analytical and numerical approaches, we present for our model conditions for 0 and $\pi$ modes to develop at the corners of the lattice. These corner modes have no static counterparts and arise from the interplay between periodic driving and many-body interactions.
\end{abstract}

\maketitle

\section{Introduction}
Ever since the seminal work by Thouless \textit{et al.} \cite{Thou1}, topological phases of matter have attracted considerable interest in the condensed matter community. One characteristic feature of such phases is the presence of robust boundary states that are insensitive to considerable perturbations and are therefore protected by nonlocal topological invariants. For the last two decades, a plethora of topological systems has been theoretically and experimentally investigated, which includes topological insulators \cite{Haldane1, KaneMele, BernevigHughesZhang, FuKane, FuKaneMele, MooreBalents, Hsieh_etal, Xia_etal, Zhang_etal1, Roy, Chen_etal,mei2014topological}, topological superconductors \cite{Kitaev1, Jiang_etal, Lutchyn_etal, Oreg_etal}, and topological semimetals \cite{Raditya1, Burkov_etal, Wan_etal, Hosur_etal, Xu_etal, PhillipsAji, Mullen_etal, Yu_etal, Kim_etal, Umer2, Umer3}. Potential technological applications of such exotic systems have been envisioned, e.g., for devising energy efficient electronic/spintronic devices \cite{Xu_etal2} and topologically-protected quantum computing \cite{Nayak_etal}.

While most existing works concern topological phases in single particle systems, many-body interaction effects are ubiquitous in nature. It has further been demonstrated that certain topological phases, such as fractional quantum Hall \cite{Laughlin1, Laughlin2, Clark_etal} and parafermionic systems \cite{Raditya3, Fendley1, AliceaFendley, Sreejith_etal, Laubscher_etal, Surace_etal, Groenendijk_etal, Thakurathi_etal, Lindner_etal, Mong_etal, Clarke_etal, ZhangKane, Orth_etal, Calzona_etal, Chew_etal}, may only exist in the presence of strong many-body interactions. For these reasons, studies of topological phases in many-body interacting systems have emerged as an active research area \cite{Rachel, Wu, XuMoore, Sirker_etal, Manmana_etal, Wang_etal1, Barbiero_etal, Roy_etal, Katsura_etal}. Interacting topological phases can generally be grouped into two different classes, i.e., symmetry protected topological (SPT) phases and topologically ordered systems \cite{Senthil}. As the name suggests, the topological classification of SPT phases, which include the Levin-Gu \cite{LevinGu}, Haldane spin-1 \cite{Haldane2}, and $\mathbb{Z}_2 \times \mathbb{Z}_2$ cluster state model \cite{Bahri_etal}, relies on the presence of some underlying symmetry. In contrast, in the topologically ordered systems, such as the Kitaev honeycomb lattice model \cite{Kitaev2}, string-net model \cite{LevinWen}, and quantum double model \cite{Kitaev3}, such a symmetry constraint is not required. \RB{It is worth noting that topologically ordered systems can further be decorated by additional symmetries to yield more exotic phases, which are usually termed symmetry enriched topological (SET) phases \cite{SET}.} 

In the efforts toward realizing topological phases, periodic driving has been identified as a powerful technique. Not only is periodic driving capable of turning an otherwise normal system into a topological one \cite{OkaAoki, McIver_etal}, but it also demonstrates the possibility of generating unique topological features with no static counterparts, such as anomalous chiral edge states with zero net Chern number \cite{Rudner_etal, Lababidi_etal, NathanRudner, ZhouGong} and boundary states pinned at half the driving frequency \cite{Raditya4, Cheng_etal, Jiang_etal, Liu_etal, Tong_etal}. It can thus be envisioned that periodically driving a many-body interacting system may give rise to rich topological phenomena. Indeed, previous studies on this subject have identified various exotic phases such as the discrete time crystals \cite{Nurwantoro_etal, Raditya5, Raditya6, Raditya7, Sacha, Else_etal1, Else_etal2, KeyserlingkSondhi, Khemani_etal, Zhang_etal2, Choi_etal, Rovny_etal1, Rovny_etal2, Pal_etal, Autti_etal, Kyprianidis_etal, Google} and parafermion modes at fractional quasienergies \cite{Raditya3}.

Motivated by these recent developments, this work presents a type of SPT phase in a two-dimensional (2D) interacting spin-1/2 lattice. Its topological feature manifests as the simultaneous existence of 0 and $\pi$ modes at the corners of the lattice. Here, 0 and $\pi$ modes respectively refer to 0 and $\omega/2$ quasienergy excitations (where $\omega$ is the driving frequency) that persist over a range of parameter values and in the presence of $\mathbb{Z}_2$ symmetry preserving perturbations. The proposed system can thus also be understood as the extension of the so-called Floquet second-order topological phases \cite{Raditya4, RadityaGong3, HuangLiu, Raditya8, Rodriguez-Vega, Hu_etal, Seshadri_etal, Plekhanov_etal, Nag_etal, PengRefael, Peng, Chaudary_etal, Zhu_etal1, Zhu_etal2, Nag1, Nag2} to the interacting setting. 

We employ three complementary approaches to probe the topology of our system. First, by fixing some parameters at specific values, we analytically derive a condition for the existence of 0 and $\pi$ modes with respect to the remaining free parameters. Here, we further establish the bulk-corner correspondence by showing that such a condition can be consistently obtained under open boundary conditions, i.e., by directly constructing both modes and checking for their convergence, as well as under periodic boundary conditions, i.e., by defining and computing appropriate topological invariants. Second, at more general parameter values we numerically evaluate the spectral functions \cite{Raditya5, Khemani_etal, Sreejith_etal, RadityaMuGong} with respect to appropriately chosen operators near a corner to capture the degeneracy and quasienergy clustering induced by the 0 and $\pi$ modes respectively. Finally, we numerically identify the simultaneous presence of 0 and $\pi$ modes from the stroboscopic time evolution profile of a spin-1/2 particle residing at a corner. In particular, the last approach also paves the way for the experimental detection of topological corner modes in our system on superconducting circuit platforms.

This paper is organized as follows. In Sec.~\ref{Sec: II}, we introduce some terminology, the present our model and briefly discuss some of its features. In Sec.~\ref{topologicalmode}, we verify the topological nature of our model by first showing analytically the emergence of 0 and $\pi$ modes at some special parameter values, and then numerically demonstrating their presence at more general parameter values. In Sec.~\ref{discussion}, we elucidate a means to detect these topological modes experimentally, as well as discuss the effect of disorder and symmetry-preserving/breaking perturbations. Finally, we summarize our work and discuss avenues for future work in Sec.~\ref{Conclusion}.

\section{Periodically driven interacting spin-1/2 lattice}
\label{Sec: II}

\subsection{Review of Floquet Theory}
\label{Sec: Floquet}

Let us briefly review the physics of periodically driven (Floquet) systems. Floquet systems are described by Hamiltonians periodic in time, i.e. $H(t+T) = H(t)$, with $T$ denoting the period. The physics of such systems is primarily governed by the one-period time evolution operator, also called the Floquet operator:
\[
    U \equiv U(T;0) = \mathcal{T}\;\text{exp}\left(-i \int_0^T H(t) dt\right).
\]
Here, we have set $\hbar = 1$. For a special class of Floquet systems where the Hamiltonian takes on the form
\begin{equation}\label{staticHamiltonians}
    H(t) = H_l \quad \text{for} \quad \frac{l-1}{N} < \frac{t}{T} \leq \frac{l}{N}
\end{equation}
where each $H_l, \;1 \leq l \leq N$ is a static Hamiltonian, $U$ is particularly easy to evaluate:
\begin{equation} \label{Floquetclass}
    U = \prod_{l=N}^{1} e^{-i H_l T/N}.
\end{equation}

Of particular interest are the system's quasienergies $\varepsilon$, derived from the Floquet eigenvalues, $e^{-i\varepsilon T}$ and the corresponding Floquet eigenstates $|\varepsilon\rangle$, i.e. $U|\varepsilon\rangle = e^{-i\varepsilon T}|\varepsilon\rangle$. These respectively play roles analogous to the energy eigenvalues and eigenstates of a usual static system. Associated with the Floquet operator $U$ is the effective Hamiltonian (also called the Floquet Hamiltonian)
\[
    H_{\text{Flo}} = iT^{-1}\;\text{ln}\; U.
\]
We can thus think of the quasienergies as eigenvalues of the Floquet Hamiltonian. Because of the logarithm function, quasienergies are only defined modulo $2\pi /T$, i.e., $|\varepsilon\rangle$ and $|\varepsilon + 2\pi n/T \rangle, \;n \in \mathbb{Z}$ represent the same state. It is customary to let $\varepsilon$ take on values in the range $(-\pi/T, \pi/T]$, which we adopt throughout this paper.

Finally, we define an $\xi$ quasienergy excitation as an operator $\gamma_\xi$ satisfying
\begin{equation} \label{excitation}
    U \gamma_\xi U^\dagger =e^{-\mathrm{i} \xi T} \gamma_\xi \;.
\end{equation}

Given any Floquet eigenstate $|\varepsilon\rangle$, $\gamma_\xi$ implies the presence of another Floquet eigenstate $|\varepsilon+\xi\rangle = \gamma_\xi |\varepsilon\rangle$, since
\begin{equation}
    U \gamma_\xi |\varepsilon\rangle = U \gamma_\xi U^\dagger U |\varepsilon\rangle = e^{-\mathrm{i} (\varepsilon+\xi) T} \gamma_\xi |\varepsilon\rangle \;.
\end{equation}

\subsection{Model}
\label{Sec: Model}

\begin{center}
\begin{figure}
    \includegraphics[scale=0.5]{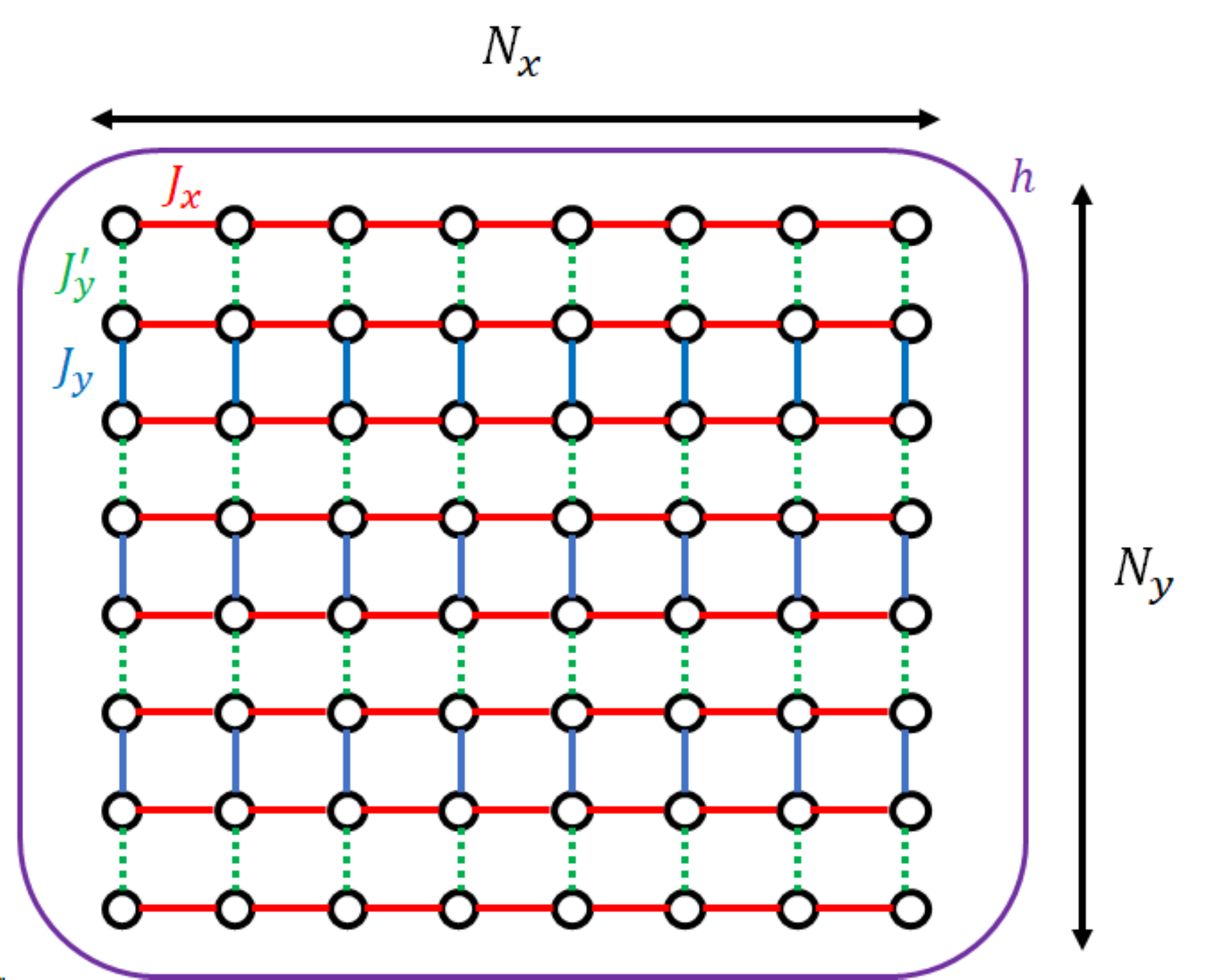}
    \caption{A schematic of the system under study. A spin-1/2 particle lives at each site of the rectangular lattice, which experiences alternate applications of an external magnetic field in the $x$-direction and nearest-neighbour $ZZ$ interactions for a $T/2$ duration.}
    \label{fig: Hamiltonian}
\end{figure}    
\end{center}

We consider a system of spin-1/2 particles arranged in a rectangular lattice and described by the time-periodic Hamiltonian
\begin{equation} \label{OurModel}
    H(t)\; = \;
    \begin{cases}
    \sum_{i,j=1}^{N_x,N_y}\;h\X_{ij}  &\text{for $nT < t \leq (n+\frac{1}{2})T$}\\
    H_x + H_y + H_y' &\text{for $(n+\frac{1}{2})T < t \leq (n+1)T$}
    \end{cases}
\end{equation}
where
\begin{equation}
\begin{split}
    H_x &= \sum_{i=1}^{N_x-1} \sum_{j=1}^{N_y} \;\quad J_x \Z_{ij}\Z_{i+1,j}\\
    H_y &= \sum_{i=1}^{N_x} \sum_{j=2,4,6\dots}^{N_y-2} J_y \Z_{ij}\Z_{i,j+1} \\
    H_y' &= \sum_{i=1}^{N_x} \sum_{j=1,3,5\dots}^{N_y-1} J_y' \Z_{ij}\Z_{i,j+1}.
\end{split}    
\end{equation}
Here $(\X/\Y/\Z)_{i,j}$ are the three Pauli Matrices at lattice site $(i,j)$, $h$ is the Zeeman field strength, $J_x$ is the $ZZ$ interaction in the $x$-direction, $J_y,J_y'$ are the $ZZ$ interactions in the $y$-direction, $T$ is the driving period, $n \in \mathbb{Z}$, and $N_x,N_y$ are the lattice sizes with $N_y$ being an even number. This system is schematically presented in Fig.~\ref{fig: Hamiltonian}.

As this Hamiltonian is of the form Eq.~(\ref{staticHamiltonians}), the associated Floquet operator is obtained via Eq.~(\ref{Floquetclass}), i.e.,
\begin{equation} 
    U=e^{-\mathrm{i} \frac{(H_x+H_y+H_y')T}{2}}\times e^{-\mathrm{i} \frac{\sum_{i,j=1}^{N_x,N_y}h \X_{i,j} T}{2}} 
    \label{Floquetmodel}
\end{equation}

\begin{center}
\begin{figure}
    \includegraphics[scale=0.5]{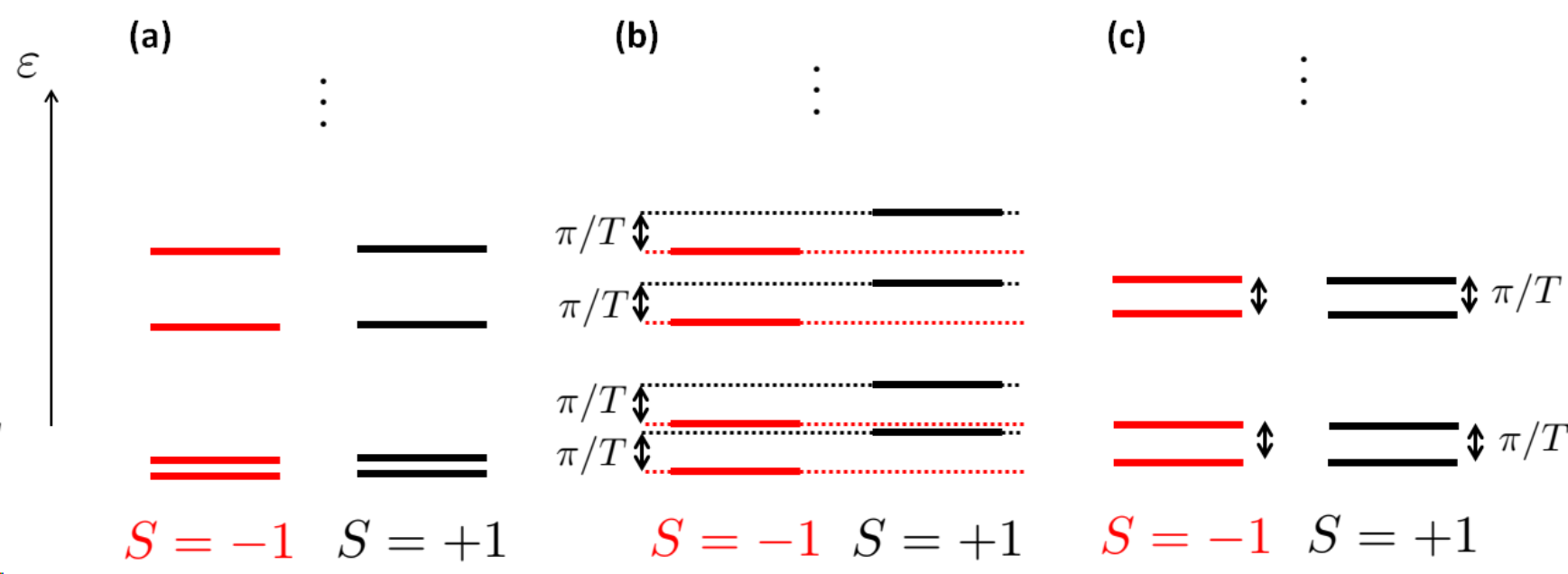}
    \caption{Typical many-body quasienergy levels in the presence of (a) ZMs, (b) PMs, and (c) both ZMs and PMs.}
    \label{fig: spectralmode}
\end{figure}    
\end{center}

Note that our Hamiltonian and its associated Floquet operator have a $\mathbb{Z}_2$ symmetry with respect to $S = \prod_{i,j}^{N_x,N_y} \X_{ij}$, i.e., $[S,H(t)]=0$ and $[S,U]=0$. Under this symmetry constraint, the system's topology can be characterized by the presence and absence of 0 and $\pi/T$ quasienergy excitations,  referred to as the 0 modes (ZMs) and $\pi$ modes (PMs) respectively.  These modes are localized at the corners. Both ZMs and PMs anticommute with $S$, which leads to a specific ordering of the even- and odd-parity quasienergy sectors of $U$. Specifically, ZMs enforce a degeneracy between the two parity sectors, whereas PMs incur a $\pi/T$ difference between pairs of quasienergies from the two parity sectors (see Fig.~\ref{fig: spectralmode}).   

\begin{figure}
    \centering
    \includegraphics[scale=0.6]{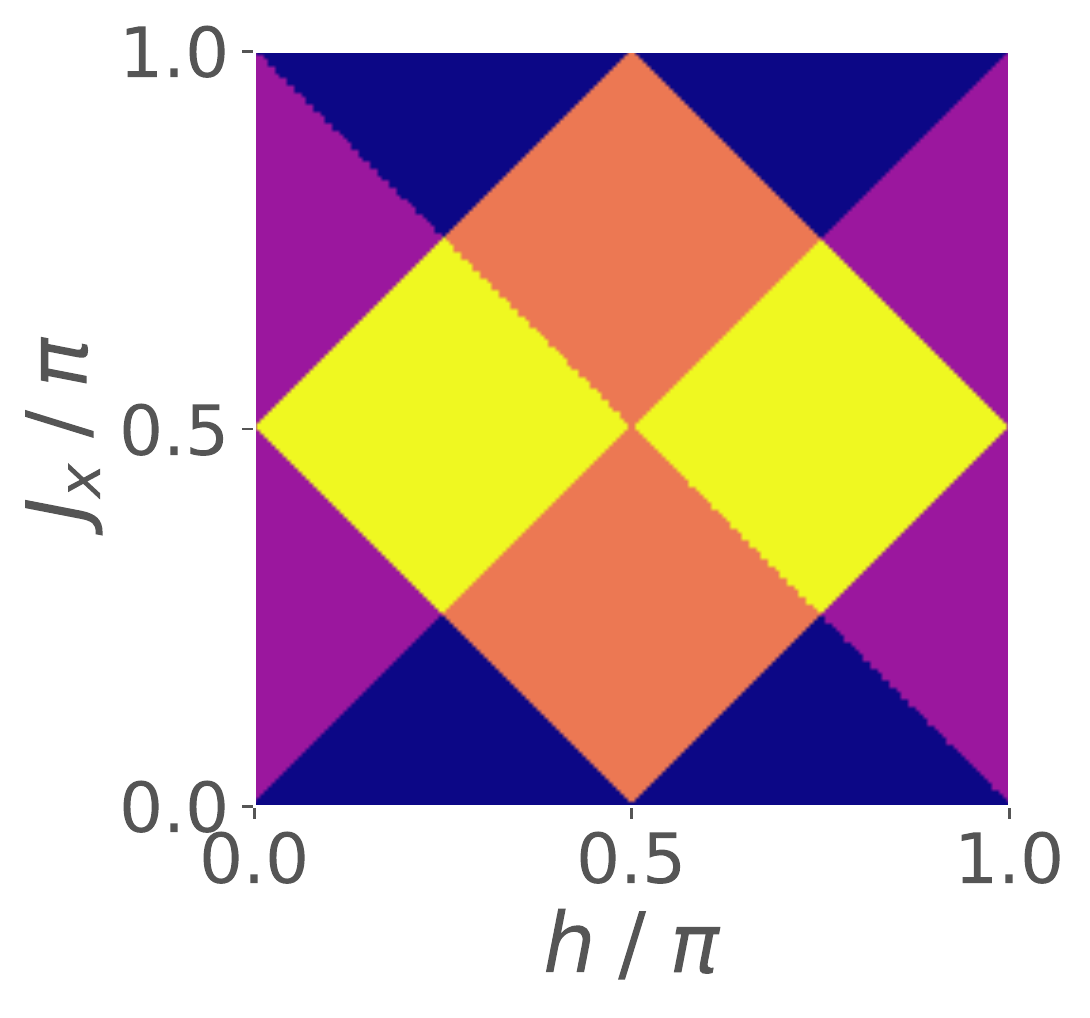}
    \caption{The phase diagram associated with Eq.~(\ref{1dmodel}). ZMs exist in the purple- and yellow-colored regime, whereas PMs exist in the peach- and yellow-colored regime. Beyond the domain $[0,1] \times [0,1]$, the diagram repeats itself.}
    \label{fig: 1D Phase Diagram}
\end{figure}

To gain insight into this system and its ability to support ZMs and PMs, we may first consider the case $J_y'=0$. It then follows from Fig.~\ref{fig: Hamiltonian} that the top and bottom chains are decoupled from the bulk akin to the flat band limit of the Su-Schrieffer-Heeger (SSH) model \cite{SSH}. Each of the boundary chains is effectively described by the one-dimensional Hamiltonian
\begin{equation}
    H(t) = 
    \begin{cases}
    \sum_{i=1}^{N} \;h\X_{i}  &\text{for $nT < t \leq (n+\frac{1}{2})T$}\\
    \sum_{i=1}^{N_x-1} J\Z_{i}\Z_{i+1}  &\text{for $(n+\frac{1}{2})T < t \leq (n+1)T$}
    \end{cases} \;
    \label{1dmodel}
\end{equation}
which support ZMs and PMs according to the well-known phase diagram in Fig.~\ref{fig: 1D Phase Diagram} \cite{Khemani_etal, Raditya3, RadityaMuGong}. In this paper, we particularly focus on $h$ and $J_x$ values within the yellow regime of Fig.~\ref{fig: 1D Phase Diagram}, where the system supports both ZMs and PMs. \RB{It should be noted that each 1D chain in the bulk is still coupled with one of its neighbors, thus resulting in the hybridization and absence of 0 and $\pi$ modes throughout the whole edge of the 2D system.} 

\RB{We now elaborate the above argument by explicitly identifying the 0 and $\pi$ modes at a system's corner and showing the absence of such modes at the system's edges.} For simplicity we shall set $T=2$ - thus the fraction $T/2$ in Eq.~(\ref{Floquetmodel}) disappears. By setting the parameter values to $h = \pi/4$ and $J_x = \pi/2$, exact expressions for ZMs and PMs can be obtained. This is achieved by finding corner-localized operators satisfying $U \gamma_0 U^\dagger = \gamma_0$ for the ZM candidates and $U \gamma_{\pi} U^\dagger = -\gamma_{\pi}$ for the PM candidates. To do so, we begin with the operators restricted to precisely a corner of the lattice, which, without loss of generality, we take to be $(1,1)$. Any such localized operator $O\in \left\lbrace \gamma_0,\gamma_\pi\right\rbrace$ can generally be written as a linear combination of the identity, and the three Pauli Matrices located at $(1,1)$. However, by taking into account symmetry considerations, i.e., $\{S, \gamma_{0}\} =  \{S, \gamma_{\pi}\} =0$, $O$ must necessarily only involve a linear combination of Pauli-Y and Pauli-Z operators. By treating the candidate equations $UOU^\dagger = \pm O$ as eigenvalue equations on the operator subspace spanned by $\{I,\X_{11},\Y_{11},\Z_{11}\}$, we then determine the appropriate coefficients required for $O$ to be a ZM or PM. Of course, this only works if the map $U(\cdot)U^{\dag}$ is closed under the set $\{I,\X_{11},\Y_{11},\Z_{11}\}$. It turns out that at the ideal parameter values $h = \pi/4, J_x = \pi/2$, this is indeed the case: $U\Y_{11}U^{\dag} = \Z_{11}$ and $U\Z_{11}U^{\dag} = \Y_{11}$. Therefore, a corner ZM and PM are respectively
\begin{equation}
\begin{split}
\gamma_{0,\rm ideal} &= \frac{1}{\sqrt{2}}\left(\Y_{11}+\Z_{11} \right)\\
\gamma_{\pi,\rm ideal} &= \frac{1}{\sqrt{2}} \left(\Y_{11}-\Z_{11} \right) \;.
\label{idealmodes}
\end{split}
\end{equation}
In a similar fashion, ZMs and PMs localized at the other three corners can be obtained. \RB{In Appendix~\ref{app1}, we repeat the above procedure with respect to operators localized at a non-corner edge site to show that ZMs and PMs are indeed absent at the system's edges.}

The above obtained corner ZMs and PMs remain present over a range of $h$ and $J_x$ values, i.e., within the yellow-colored regime in Fig.~\ref{fig: 1D Phase Diagram}. In the following section, we will demonstrate that such ZMs and PMs survive at nonzero $J_y'$ values as well, thus establishing their topological nature in the 2D system.

\section{Topological corner ZMs and PMs}
\label{topologicalmode}
\textcolor{black}{Since $J_y'$ represents an intra-cell coupling in the $y$-direction, understanding its interplay with the corresponding inter-cell coupling $J_y$ in the same direction is insightful. It is therefore instructive to investigate the fate of these ZMs and PMs as $J_y'$ deviates from $0$.} At $J_y'\neq 0$, all horizontal chains are coupled with one another. As a result, the eigenvalue equation $U\gamma_\xi U^\dagger =e^{-\mathrm{i}\xi T} \gamma_\xi$ becomes a genuine 2D many-body problem that is generally analytically intractable even at moderate system sizes. To establish their topological nature, it is however necessary to verify that ZMs and PMs remain present over a range of parameter values. To tackle this more general scenario, we employ the following two strategies. 

In Sec.~\ref{analytic}, we first focus on the parameter values $h = \pi/4$ and $J_x = \pi/2$ to gain enough simplification which allows for the explicit identification of ZMs and PMs, as well as the condition for their existence. At these parameter values, appropriate topological invariants can further be defined and analytically computed to establish the bulk-corner correspondence. In Sec.~\ref{numeric}, we then consider more general parameter values that warrant full numerical treatment for the characterization of ZMs and PMs. 

\subsection{Corner modes at $h = \pi/4$ and $J_x = \pi/2$}
\label{analytic}

\subsubsection{Corner modes from eigenvalue equation on operator space}

\begin{figure}
    \centering
    \vspace{8pt}
    \includegraphics[scale=0.6]{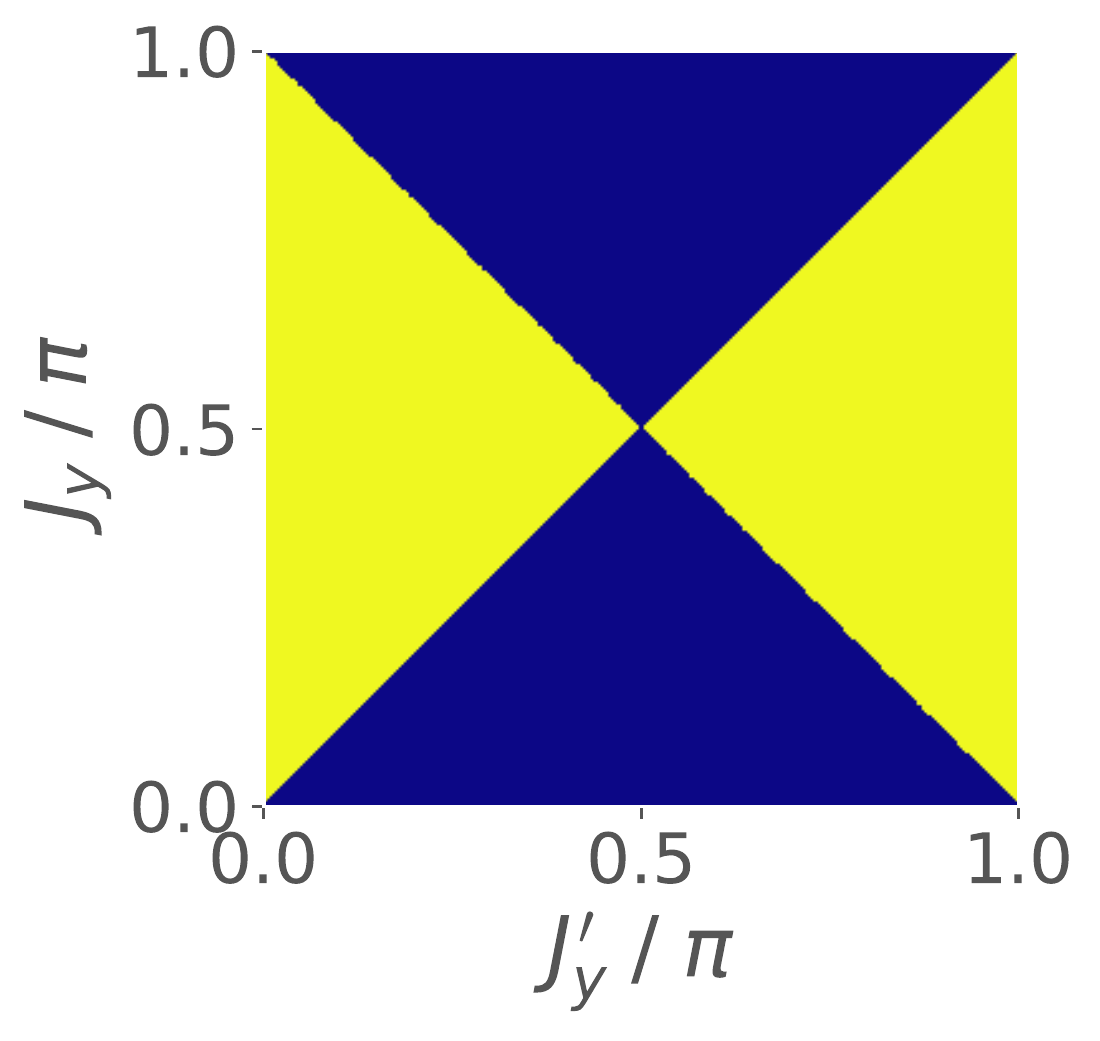}
    \caption{The phase diagram derived analytically from our model, Eq.\;(\ref{OurModel}). This is a graph of ${\rm sgn}(\text{cos}\; 2J_y' - \text{cos}\; 2J_y)$, which outputs 1 (yellow) when $\text{cos}\; 2J_y < \text{cos}\; 2J_y'$ and 0 (blue) otherwise. Both ZMs and PMs exist in the yellow-colored regime, and do not exist in the blue-colored regime.}
    \label{fig: Main Phase Diagram}
\end{figure}

With the specific parameter values given above, it can be verified that the Floquet operator maps a set of Pauli matrices $\left\lbrace \prod_{j=1}^{N_y} (\sigma_{i,j}^{x})^{p_j}(\sigma_{i,j}^{y})^{q_j}: p_j,q_j\in\left\lbrace 0,1\right\rbrace \right\rbrace$ into itself, i.e., any vertical chain is effectively decoupled from its neighbors. To find explicitly the ZMs and PMs, we generalize the approach outlined at the end of Sec.~\ref{Sec: Model}, the details of which are presented in Appendix \ref{app1}. We then find that both ZMs and PMs exist (i.e. normalizable, and therefore physical) if the condition $\text{cos} \;2J_y < \text{cos} \;2J_y'$ is satisfied; otherwise, neither ZMs nor PMs are present. Our result is summarized in the phase diagram of Fig.~\ref{fig: Main Phase Diagram}.   

\subsubsection{Corner modes from topological invariant}

The above criterion for the existence of corner modes can alternatively be reproduced by defining and computing appropriate topological invariants. To this end, let $\tilde{U}$ be the Floquet operator acting on the projection of the system onto the left, vertical boundary. We find that it can be represented as
\begin{eqnarray}
\tilde{U} &=& e^{-\mathrm{i} \left(\sum_{j=1}^{N/2-1} J_y \sigma_{2j}^z \sigma_{2j+1}^z +\sum_{j=1}^{N/2} [J_y'+\frac{\pi}{2}] \sigma_{2j-1}^z \sigma_{2j}^z \right) } \nonumber \\
&& \times e^{-\mathrm{i} \frac{\pi}{4} \sum_{j=1}^{N}\sigma_j^x} \;,
\end{eqnarray}
where for notational simplicity,  we have suppressed the (first) index referencing the horizontal position, which is always $1$, and written $N_y=N$. Indeed, it can be directly verified that $\tilde{U}P\tilde{U}^\dagger$ and $UPU^\dagger$ yield identical results for any $P$ being a product of Pauli operators residing at the leftmost chain. Importantly, $\tilde{U}$ only has support strictly on the leftmost chain, i.e., it is effectively 1D. We then construct the corresponding edge Hamiltonian that generates $\tilde{U}$ over one period. By applying a Jordan-Wigner transformation with respect to such an operator, we obtain an effective 1D fermionic Hamiltonian
\begin{equation}
    H_{\text{eff}}(t)\; = \;
    \begin{cases}
     H_1 &\text{ for } 2\ell<t<2\ell+1\\
     H_2 &\text{ for } 2\ell+1<t<2\ell+2,
    \end{cases}
\end{equation}
where 
\begin{equation}
\begin{split}
H_1 &= \sum_{j=1}^{N}\;\frac{\pi}{4} \left(2a_i^{\dag}a_i-1 \right) \;, \\
H_2 &= \sum_{j=1\dots}^{N/2-1} \; J_y \left(a_{2i}^{\dag}a_{2i+1} + a_{2i}^{\dag}a_{2i+1}^{\dag} + h.c \right)\\
&+ \sum_{j=1\dots}^{N/2} \; (J_y'+\frac{\pi}{2}) \left(a_{2i-1}^{\dag}a_{2i} + a_{2i-1}^{\dag}a_{2i}^{\dag}  + h.c. \right)\;,
\end{split}
\end{equation}
$\ell\in \mathbb{Z}^+$, and $a_i^\dag$ is a fermionic creation operator.
It is now helpful to view our system as a lattice of composite cells, each cell comprising sublattice sites A and B, with intracell coupling $J_y'+\pi/2$ and intercell coupling $J_y$, just like in the SSH Model. Relabelling terms gives
\begin{equation}
\begin{split}
    H_1 &=\sum_{j=1}^{N}\;\frac{\pi}{4} \left(2\left(a_{i,A}^{\dag}a_{i,A} + a_{i,B}^{\dag}a_{i,B}\right)-2 \right) \;,\\
    H_2 &= \sum_{j=1}^{N-1} \; J_y \left(a_{i,B}^{\dag}a_{i+1,A}^{\dag} + a_{i,B}^{\dag}a_{i+1,A} + h.c. \right)\\
    &+ \sum_{j=1}^{N} (J_y'+\frac{\pi}{2}) \left(a_{i,A}^{\dag}a_{i,B}^{\dag} + a_{i,A}^{\dag}a_{i,B} + h.c. \right) ,
\end{split}
\end{equation}
where we have redefined $N/2\rightarrow N$. Under periodic boundary conditions, we apply the Fourier Transform $a_{j,A/B}^{\dag} = \frac{1}{\sqrt N} \sum_{k \in BZ} e^{-ijk}d_{k,A/B}^{\dag}$ (thus the momentum numbers take on the values $k=\frac{2\pi n}{N}, n \in [-\frac{N}{2},\frac{N}{2}]\cap \mathbb{Z}$), and make some rearrangements among the terms to enable the Bogoliubov-de Gennes description
\begin{equation}
    H_{\text{eff}}(t) = \frac{1}{2} \sum_{k \in BZ} \Psi_k^{\dag}H_{\text{BdG}}(k,t)\Psi_k
\end{equation}
where $\Psi_k^{\dag} = [d_{k,A}^{\dag}, d_{k,B}^{\dag}, d_{-k,A}, d_{-k,B}]$. $H_{\text{BdG}}(k,t)$ alternates between $H_{1,\text{BdG}}(k)$ and $H_{2,\text{BdG}}(k)$ every $T/2$ units of time, where
\begin{equation}
\begin{split}
    H_{1,\text{BdG}}(k) = \;&\frac{\pi}{2} \Z \otimes I\\
    H_{2,\text{BdG}}(k) = \;&\left(J_y \text{cos}\;k + (J_y'+\frac{\pi}{2})\right) \Z \otimes \X  \\
    & +\left(J_y \text{cos}\;k - (J_y'+\frac{\pi}{2})\right) \Y \otimes \Y  \\
    & + \left(J_y \text{sin}\;k \right) \Z \otimes \Y - \left(J_y \text{sin}\;k \right) \Y \otimes \X . 
\end{split}
\end{equation}

Note that both $H_{1,\text{BdG}}(k)$ and $H_{2,\text{BdG}}(k)$ respect the chiral symmetry under $\mathcal{C}=\sigma_x\otimes I$, i.e., $\mathcal{C}H_{1,\text{BdG}}(k)\mathcal{C}^\dagger=-H_{1,\text{BdG}}(k)$ and $\mathcal{C}H_{2,\text{BdG}}(k)\mathcal{C}^\dagger=-H_{2,\text{BdG}}(k)$. We can then follow the prescription of Ref.~\cite{Asboth_etal} to define winding number invariants $\nu_0$ and $\nu_\pi$ that respectively count the number of 0 modes and $\pi$ modes at each edge of the 1D chain. To this end, we first write the symmetric-timeframe Floquet operator (i.e. the Floquet operator shifted in time by $T/4$) associated with $H_{\text{BdG}}(k,t)$ in the form
\begin{equation}
    \mathcal{U}_{\rm BdG}(k) = G(k) F(k) \;,
\end{equation}
where 
\begin{equation}
F(k)=\mathcal{C}^\dagger G(k)^\dagger \mathcal{C} = e^{-\mathrm{i} H_{2,\text{BdG}}(k)/2} \cdot e^{-\mathrm{i} H_{1,\text{BdG}}(k)/2} \;.    
\end{equation}
In the canonical basis where $\mathcal{C}$ is diagonal, $F$ can be written in matrix form as
\begin{equation}
    F(k)=\left[\begin{array}{cc}
        A(k) & B(k) \\
        C(k) & D(k) 
    \end{array}\right] \;,
\end{equation}
where $A(k)$, $B(k)$, $C(k)$, and $D(k)$ are $2\times 2$ block matrices. The exact expressions of $A(k)$ and $C(k)$ are unimportant, whereas those of $B(k)$ and $D(k)$ are presented in Appendix~\ref{app2}. Following Ref.~\cite{Asboth_etal}, we can then find $\nu_0=|\nu[B]|$ and $\nu_\pi=|\nu[D]|$, where
\begin{equation}
        \nu[M] = \frac{1}{2\pi i} \int_{-\pi}^{\pi} dk \frac{d}{dk} \text{ln} \;[\text{det}\;M(k)]
\end{equation}
is the winding number. By analytically evaluating these winding numbers (see Appendix~\ref{app2} for details), we find that $\nu_0=\nu_\pi=1$ if and only if $\cos 2J_y <\cos 2J_y'$. Going back to the full $2D$ picture, we thus conclude that, with the special $h$ and $J_x$ values under consideration, the corner ZMs and PMs simultaneously exist if and only if the vertical interaction parameters $J_y$ and $J_y'$ satisfy $\cos 2J_y <\cos 2J_y'$. 

Note that this is precisely the same conclusion we arrived at through our first analysis, namely the examination of an eigenvalue equation on operator space. This highlights the topological nature of the corner ZMs and PMs. 

%\textbf{Result}
%\label{MainResult}
%\begin{tcolorbox}[ams align*]
%\text{cos} \;2J_y < \text{cos} \;2J_y' &\implies \text{Both modes exist}\\
%\text{cos} \;2J_y > \text{cos} \;2J_y' &\implies \text{No modes exist}    
%\end{tcolorbox}

\subsection{Corner modes at general parameter values}
\label{numeric}

\textcolor{black}{
To determine the fate of ZMs and PMs for more general values (but still within the yellow-colored regime in Fig.~\ref{fig: 1D Phase Diagram}) of $h$ and $J_x$, we now resort to numerics. To this end, we define spectral functions around the 0 and $\pi/T$ quasienergies:
\begin{equation}
\begin{split}
    s_0 &= \frac{1}{\mathcal{N}_0} \sum_{\ket{\varepsilon,i} \in \chi} \int_{-\Delta}^{\Delta} S_0(\ket{\varepsilon,i}, \eta) \;d\eta\\
    s_{\pi} &= \frac{1}{\mathcal{N}_{\pi}} \sum_{\ket{\varepsilon,i} \in \chi} 
    \int_{\pi-\Delta}^{\pi+\Delta} S_{\pi}(\ket{\varepsilon,i}, \eta) \;d\eta.
    \label{specfun}
\end{split}
\end{equation}
Here $\Delta$ is a small real number \textcolor{black}{(we have used $\Delta=0.05$ throughout)}, $\mathcal{N}_{0/\pi} = \sum_{\ket{\varepsilon,i} \in \chi} \int_{-\pi}^{\pi} S_{0/\pi}(\ket{\varepsilon,i},\eta) \;d\eta$ serve as normalization factors, and the integrands are
\begin{equation}
\begin{split}
    S_0(\ket{\varepsilon,i},\eta) &= \sum_{\ket{\varepsilon',j} \in X} \delta(\varepsilon - \varepsilon' - \eta)|\bra{\varepsilon',j}\gamma_0\ket{\varepsilon,i}|^2\\
    S_{\pi}(\ket{\varepsilon,i},\eta) &= \sum_{\ket{\varepsilon',j} \in X} \delta(\varepsilon - \varepsilon' - \eta)|\bra{\varepsilon',j}\gamma_{\pi}\ket{\varepsilon,i}|^2.
    \label{s0spi}
\end{split}
\end{equation}
Let us first clarify our notation. $\ket{\varepsilon,i}$ refers to an eigenstate in the eigenspace associated with the quasienergy $\varepsilon$, and $i$ indexes the degeneracy of this eigenspace. $X$ is the set of all Floquet eigenstates, and $\chi \subset X$ is a subset thereof. %The notation $\sum_{\ket{\varepsilon,i} \in S}$ means sum over all eigenstates in the set $X/\chi$. These eigenstates to be summed over are generically denoted $\ket{\varepsilon,i}$, although they could of course be of different quasienergies.
}

\textcolor{black}{
Intuitively, $S_0(\ket{\varepsilon,i},\eta)$ [$S_\pi(\ket{\varepsilon,i},\eta)$] quantifies the goodness of $\gamma_0$ [$\gamma_\pi$] as a ZM [PM]. The Dirac delta function appearing in Eq.~(\ref{s0spi}) ensures that $S_0(\ket{\varepsilon,i},\eta)$ [$S_\pi(\ket{\varepsilon,i},\eta)$] is only contributed by $\bra{\varepsilon',j}\gamma_0\ket{\varepsilon,i}|^2$ [$\bra{\varepsilon',j}\gamma_{\pi}\ket{\varepsilon,i}|^2$] involving two quasienergies that are sufficiently close to each other [close to having $\pi$ separation]. Indeed, if $\gamma_0$ is close to being the actual ZM (if it exists), then $\gamma_0\ket{\varepsilon,i} = \ket{\varepsilon',j}$ for some $\varepsilon' \approx \varepsilon$. The integration of $S_0(\ket{\varepsilon,i},\eta)$ over the small window $[-\Delta,\Delta]$ then gives approximately $1$. However, if $\gamma_0$ is nowhere close to being a ZM, then $\gamma_0\ket{\varepsilon,i}$ is generally `scattered' throughout the entire Hilbert Space, and has considerable support on $\varepsilon'$-eigenspaces with $\varepsilon'$ being far from $\varepsilon$ ($|\varepsilon'-\varepsilon|>\Delta$). Here, the Dirac delta function filters out all such components, so integration over $[-\Delta,\Delta]$ gives approximately $0$.
}

\textcolor{black}{Note that even if $\gamma_0$ is nowhere close to being a ZM, it is still possible, though very unlikely, that certain eigenstates $\ket{\varepsilon,i}$ coincidentally give $\gamma_0\ket{\varepsilon,i} = \ket{\varepsilon',j}$ with $\varepsilon' \approx \varepsilon$. The summation over $\chi$ in Eq.~(\ref{specfun}) is introduced to avoid such a false positive. Ideally, taking $\chi$ to be the set of all Floquet eigenstates gives the best result. However, doing so not only becomes impractical at moderate to large system sizes, but it is also unnecessary to yield a convincing result. In particular, as long as $\chi$ samples a considerable number of Floquet eigenstates, the false positive contribution due to the above event will be negligibly small. In our investigations, we chose $|\chi|=30$ out of the full set of size $|X|= 2^{N_xN_y}$. (c.f. $|\chi|=8$ in Ref.~\cite{Sreejith_etal}).}

%\RB{Thus we cannot simply
%evaluate $S_0(\ket{\varepsilon,i},\eta)$ for a single $\ket{\varepsilon,i}$, for the very reason above that we might get unlucky. In theory we should evaluate $S_0(\ket{\varepsilon,i},\eta)$ for all $\ket{\varepsilon,i} \in X$ then take the average to give $s_0$, but in practice a small, randomly chosen subset $\chi$ more than suffices. This is because probability compounds, and it becomes \textit{extremely} unlikely to have $\gamma_0$ being not (close to) a ZM, yet $\gamma_0\ket{\varepsilon,i} = \ket{\varepsilon',j}$ and $\varepsilon' \approx \varepsilon$ for many $\ket{\varepsilon,i}$. In our investigations, we chose $|\chi|=30$, out of the full set of size $2^{N_xN_y}$. (See also \cite{Sreejith_etal}, where $|\chi|=8$.) Thus in the equation for the average, $s_0$, the sum is over $\chi$, not $X$.
%}

\textcolor{black}{The above reasoning applies, mutatis mutandis, to the PM case, with $\gamma_0,S_0,s_0 \longrightarrow \gamma_{\pi},S_{\pi},s_{\pi}$, and $(\varepsilon' \approx \varepsilon) \longrightarrow (\varepsilon' \approx \varepsilon+\pi)$. To summarize, $s_0$ and $s_\pi$ precisely quantify the tendency of the system to exhibit the spectral properties of Fig.~\ref{fig: spectralmode} under $\gamma_0$ and $\gamma_\pi$ respectively. That is, $s_0=1$ ($s_\pi=1$) if a ZM (PM) exists \emph{and} is exactly represented by the operator $\gamma_0$ ($\gamma_\pi$). 
}

%Intuitively, $S_0(\ket{\varepsilon,i},\eta)$ [$S_\pi(\ket{\varepsilon,i},\eta)$] quantifies the tendency of the quasienergy $\varepsilon$ to display a two-fold degeneracy [$\pi/T$ spacing] induced by the operator $\gamma_0$ [$\gamma_\pi$], up to a small uncertainty $\Delta$.

For general parameter values, the exact expressions of $\gamma_0$ and $\gamma_\pi$, if they exist, are not readily available in closed forms. As such, we will be using Eq.~(\ref{idealmodes}) for the evaluation of the above spectral functions. In particular, if ZMs and PMs remain present at a given set of parameter values, they are localized at a corner and are expected to have considerable overlap with those in Eq.~(\ref{idealmodes}). Therefore, parameter regimes which support ZMs (PMs) are identified by a finite, but less than unity, $s_0$ ($s_\pi$) value.

\RB{Since our system is characterized by four parameters, evaluating the full dependence of the spectral functions on $h$, $J_x$, $J_y$, and $J_y'$ is very challenging. In the following, we choose to focus on varying $J_y'$ and $J_y$. This amounts to investigating the robustness of our analytically obtained corner ZMs and PMs against the presence of coupling between the top/bottom chain and the rest of the bulk. While one may, in principle, also consider the spectral functions at varying $J_y'$ and $J_x$, we choose not to pursue it in this paper due to the limitations of our computer. Specifically, even if both ZMs and PMs remain present at a given set of parameter values, their localization length in the $x$-direction ($y$-direction) depends on $J_x$ ($J_y'$ and $J_y$). Therefore, to accurately capture the presence of ZMs and PMs at varying $J_y'$ and $J_x$, a sufficiently large $N_x$ and $N_y$ is necessary. In contrast, to achieve the same at varying $J_y'$ and $J_y$, only $N_y$ is required to be large. Nevertheless, to ensure that our choice of parameters is as nontrivial as possible, we have chosen $J_x$ and $h$ to deviate from their analytically solvable values in our numerics below.}

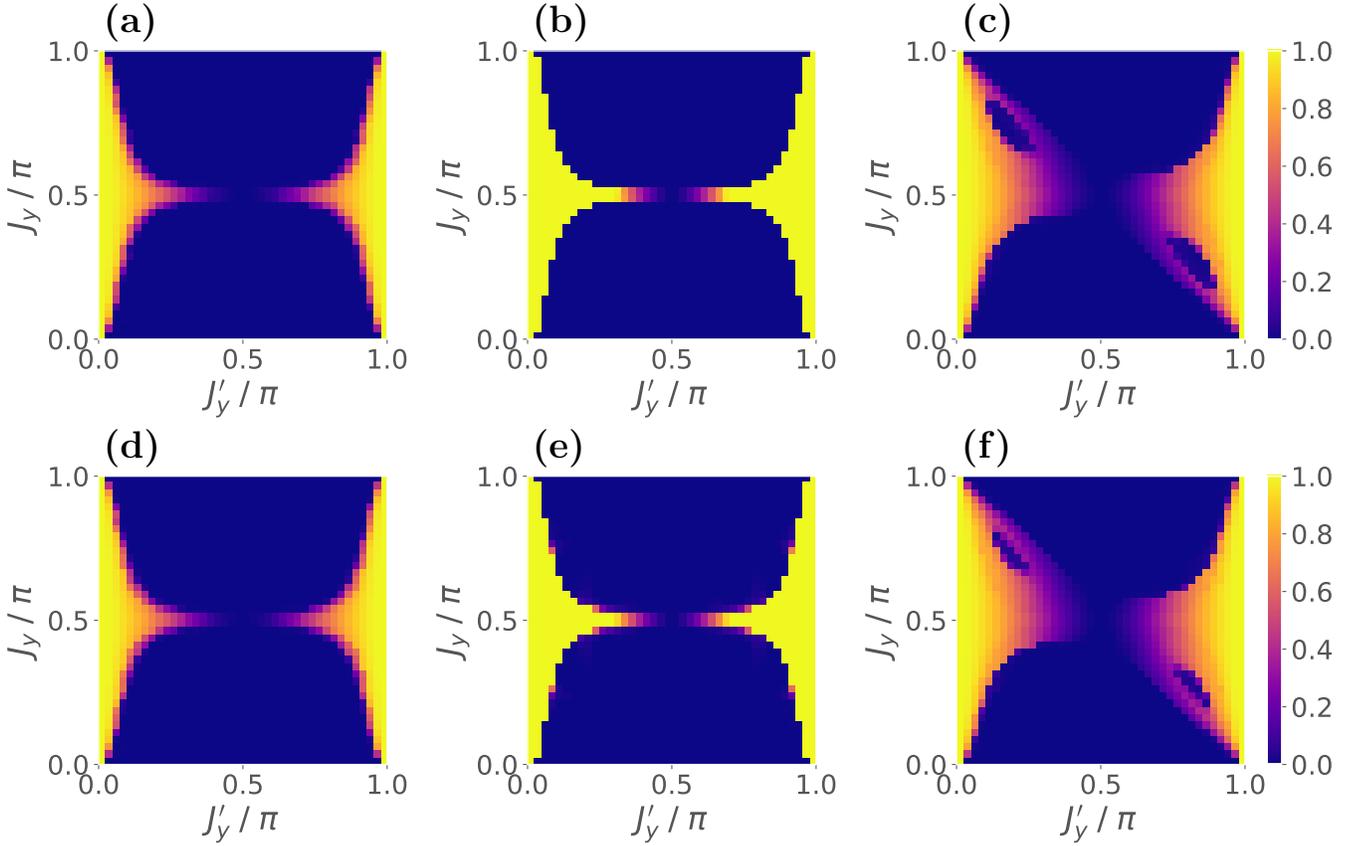
\begin{figure*}
	\centering
	\begin{minipage}{0.3\linewidth}
		\begin{picture}(160,160)
			\put(-10,0){\includegraphics[scale = 0.5]{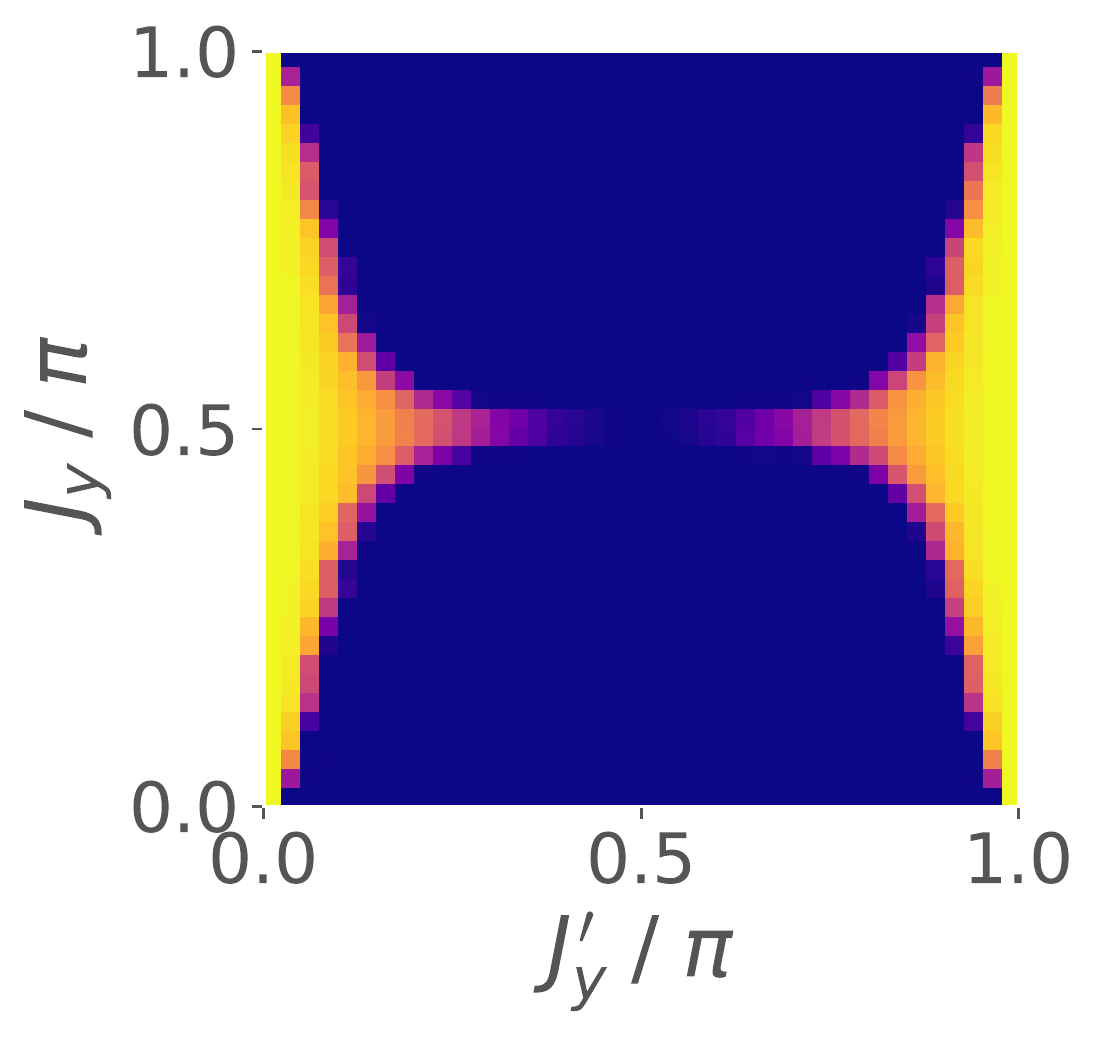}}		
			\put(30,150){\textbf{\Large{(a)}}}	
		\end{picture}
	\end{minipage}
	~
	\begin{minipage}{0.3\linewidth}
		\begin{picture}(160,160)
			\put(-10,0){\includegraphics[scale = 0.5]{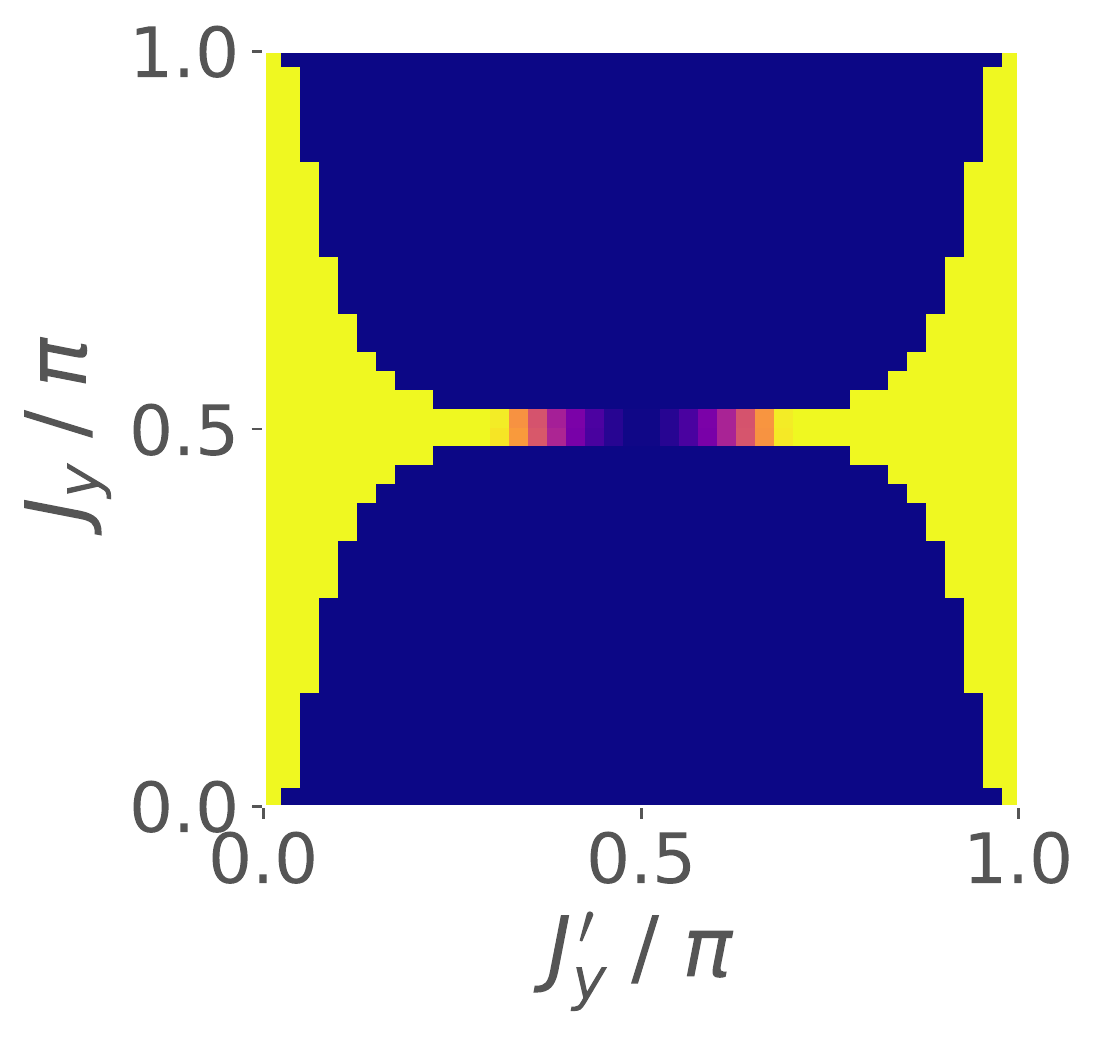}}	
			\put(30,150){\textbf{\Large{(b)}}}	
		\end{picture}
	\end{minipage}
    ~
	\begin{minipage}{0.3\linewidth}
		\begin{picture}(160,160)
			\put(-10,0){\includegraphics[scale = 0.5]{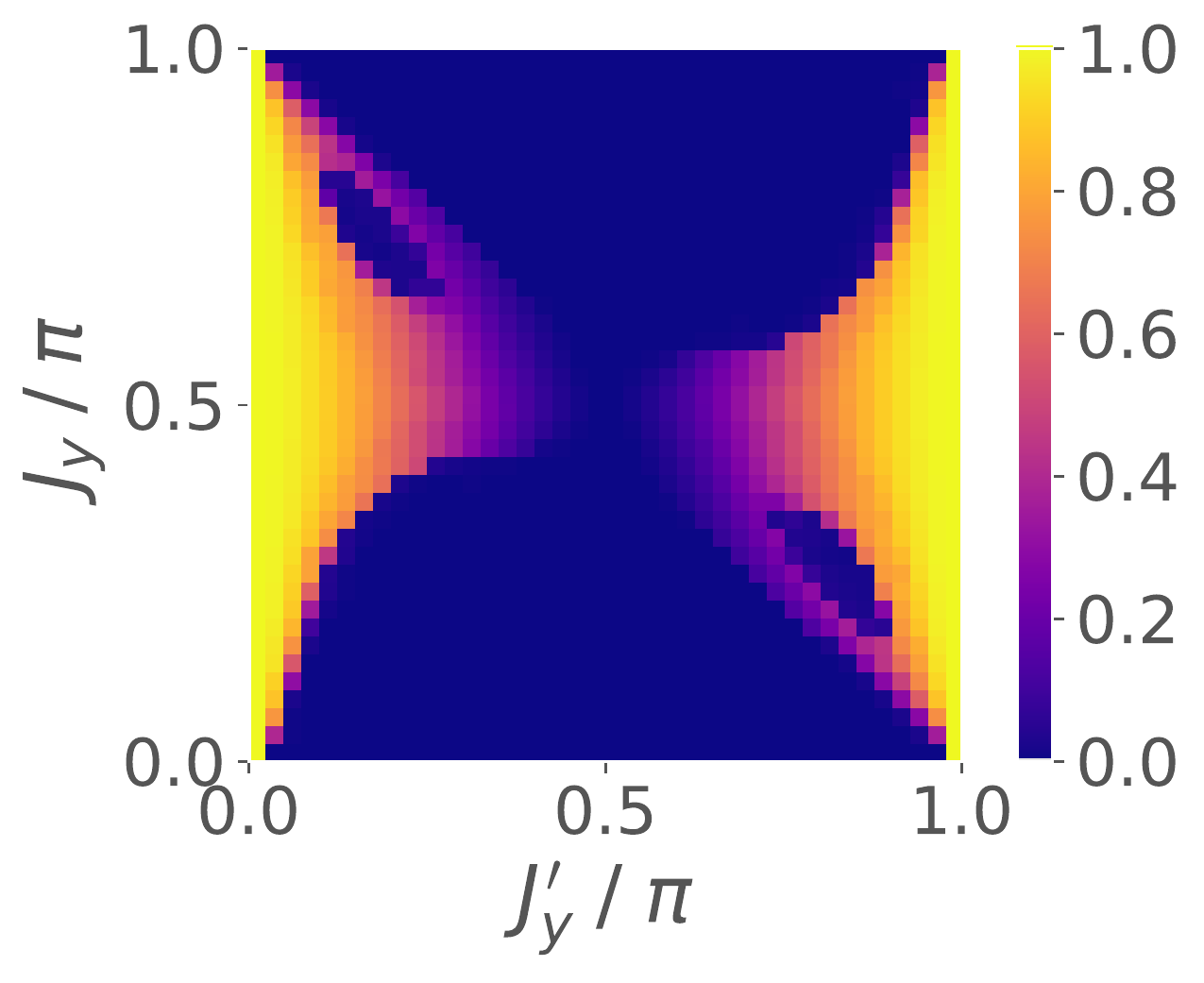}}		
			\put(30,150){\textbf{\Large{(c)}}}	
		\end{picture}
	\end{minipage}
	\\  
	\begin{minipage}{0.3\linewidth}
		\begin{picture}(160,160)
			\put(-10,0){\includegraphics[scale = 0.5]{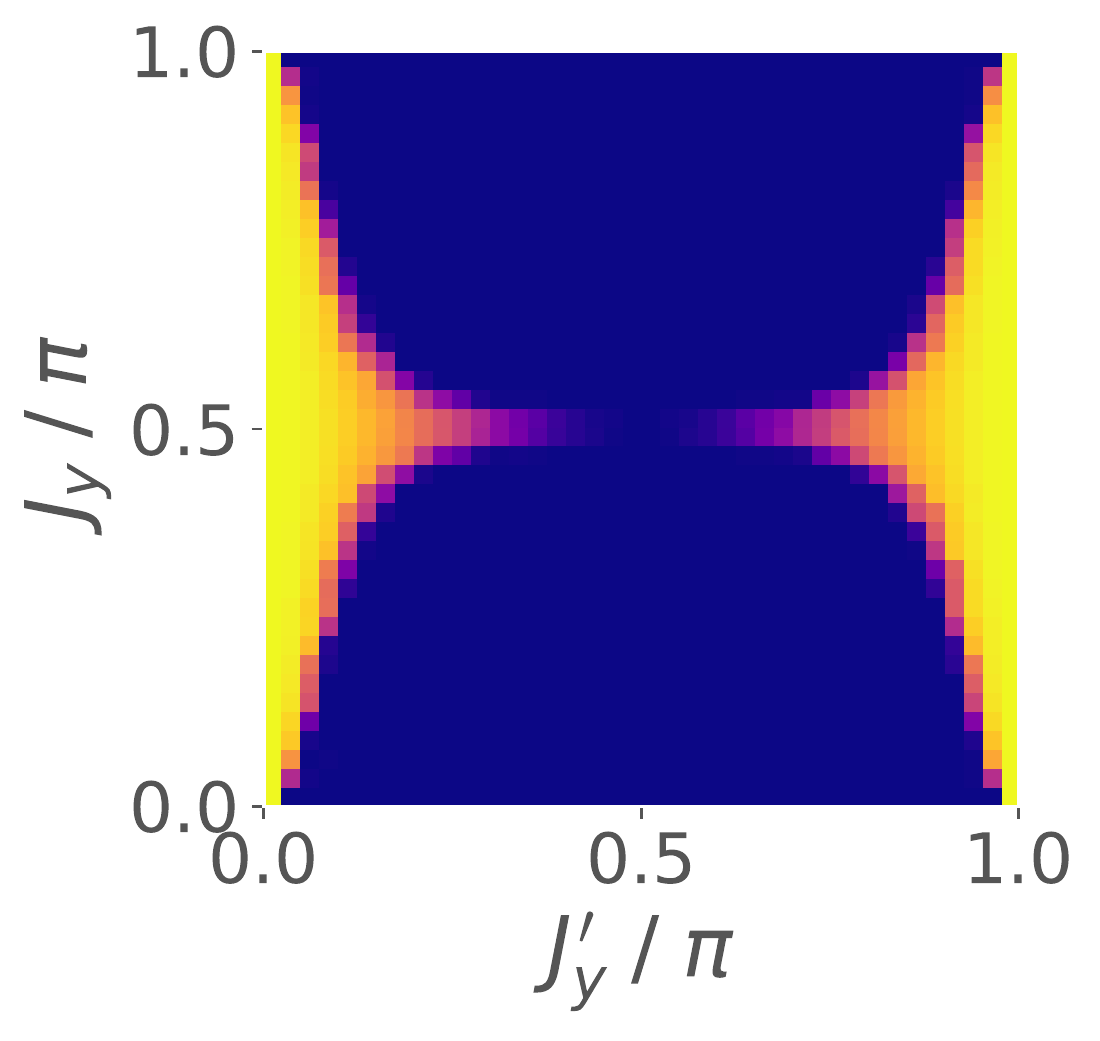}}		
			\put(30,150){\textbf{\Large{(d)}}}	
		\end{picture}
	\end{minipage}
	~
	\begin{minipage}{0.3\linewidth}
		\begin{picture}(160,160)
			\put(-10,0){\includegraphics[scale = 0.5]{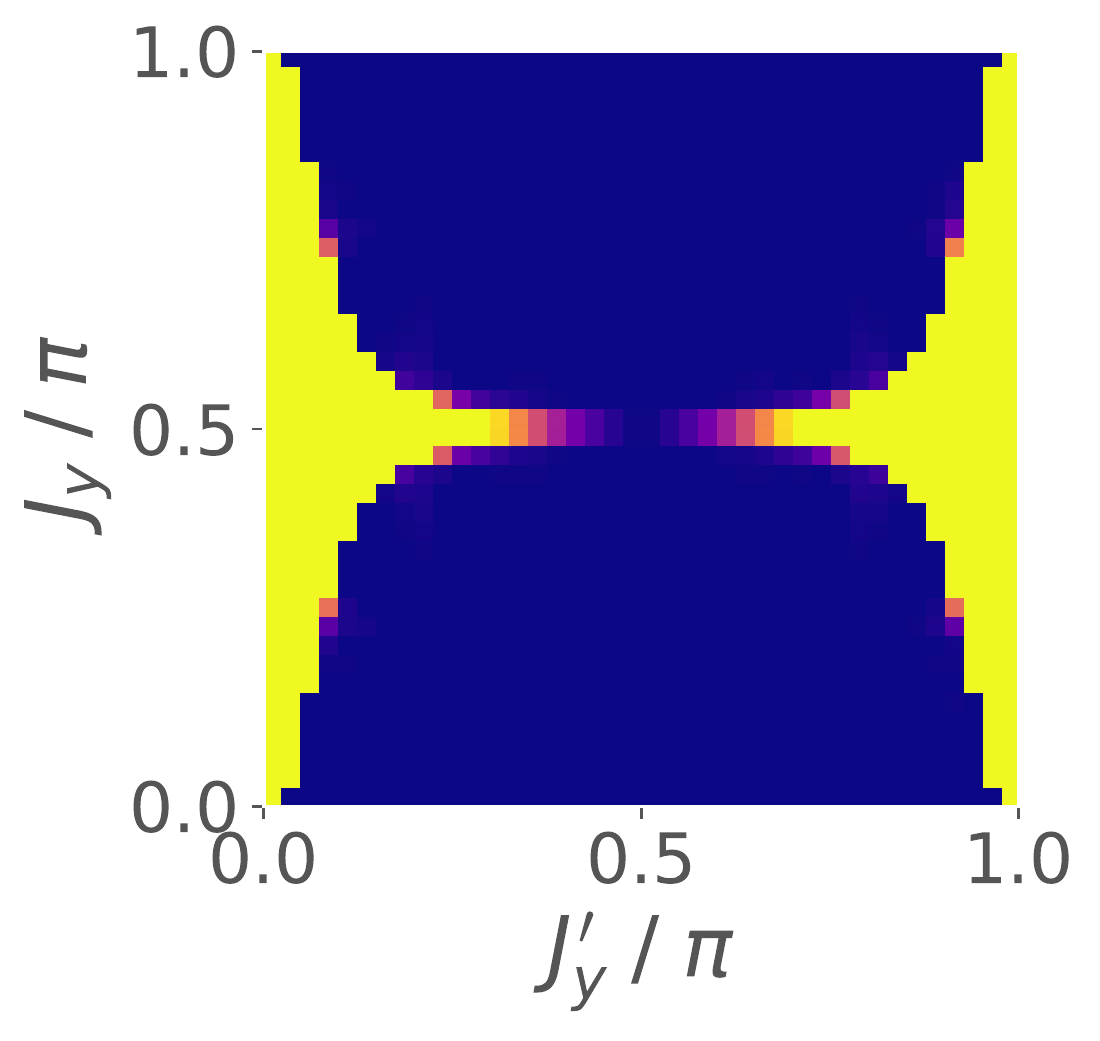}}
			\put(30,150){\textbf{\Large{(e)}}}	
		\end{picture}
	\end{minipage}
    ~
	\begin{minipage}{0.3\linewidth}
		\begin{picture}(160,160)
			\put(-10,0){\includegraphics[scale = 0.5]{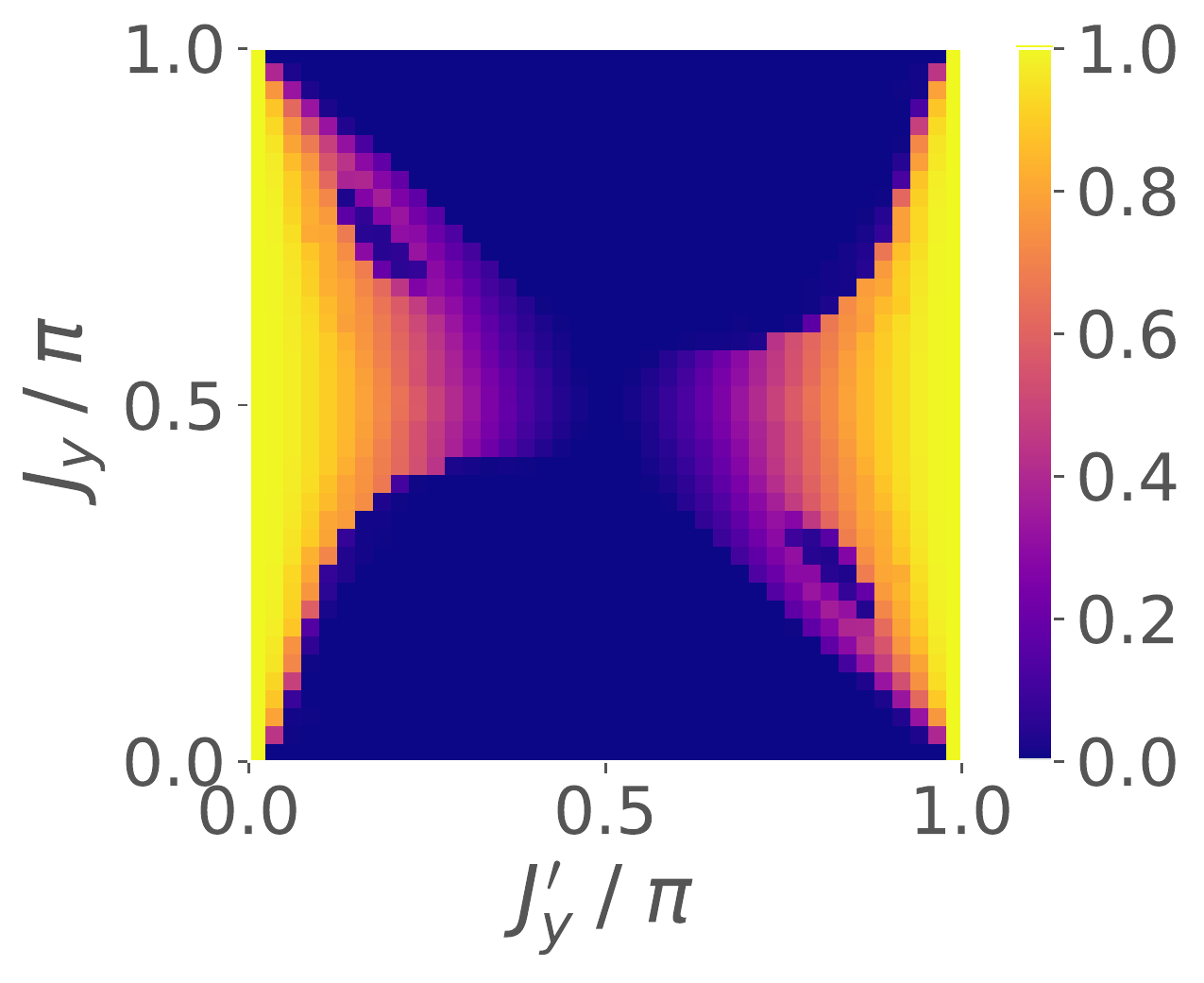}}
			\put(30,150){\textbf{\Large{(f)}}}	
		\end{picture}
	\end{minipage}
	
	\caption{\textcolor{black}{
	Spectral functions for (a,b,c) ZMs ($s_0$) and (d,e,f) PMs ($s_{\pi}$). The system size is chosen as (a,d) $2 \times 4$, (b,e) $3 \times 4$, and (c,f) $2 \times 6$. These spectral functions are evaluated over $(J_y,J_y') \in [0,\pi] \times [0,\pi]$, while the other system parameter values are fixed at $h=1.01\pi/4$ and $J_x=1.01\pi/2$.
	}}
    \label{fig: Spectral Function diagrams}
\end{figure*}

In Fig.~\ref{fig: Spectral Function diagrams}, we present the numerically calculated spectral functions for a wide range of parameter values $J_y$ and $J_y'$. The remaining parameters $h$ and $J_x$ are fixed at values within the yellow regime in Fig.~\ref{fig: 1D Phase Diagram}, but (slightly) away from the analytically solvable point, i.e. we set $(h,J_x)=(1.01\pi/4,1.01\pi/2)$. \RB{Due to the computer limitations for the exact diagonalization of quantum many-body systems and the constant need to exponentiate the Hamiltonians, only three system sizes are considered in Fig.~\ref{fig: Spectral Function diagrams}: $2\times4$, $3\times4$ and $2\times6$. However, even with these system sizes, it is already clearly observed that the increase in $N_y$ leads to a profile closer to the analytically obtained phase diagram of Fig.~\ref{fig: Main Phase Diagram}. Moreover, consistent with our previous analytical results, we observe that $s_0\approx s_\pi$.} 

\RB{Finally, we also see that increasing $N_x$ while fixing $N_y$ ($2\times4$ vs $3\times4$) does not qualitatively affect the profiles of the spectral functions appreciably. This is due to the ZMs and PMs having a sufficiently small localization length along the $x$-direction under the parameters $h$ and $J_x$ we considered. In this case, hybridization between modes near the left and right corners due to their overlap is already negligibly small even at the smallest system size $N_x$ we considered. As a result, increasing $N_x$ does not yield further improvement.}

\section{Discussion}
\label{discussion}

The Floquet operator associated with the two-time-step Hamiltonian studied in this paper can be naturally implemented with superconducting qubits. To this end, it is instructive to first rewrite Eq.~(\ref{Floquetmodel}) as
\begin{equation}
    U=\left(\prod_{i,j} U_{x,i,j} U_{y,i,j} \right) \left(\prod_{i,j} U_{h,i,j} \right) \;,  
    \label{Floquetcircuit}
\end{equation}
where
\begin{equation}
\begin{split}
    U_{x,i,j} &= e^{-\mathrm{i} J_x \sigma_{i,j}^z \sigma_{i+1,j}^z}\\
    U_{y,i,j} &= e^{-\mathrm{i} \frac{J_y+J_y' +(-1)^j(J_y-J_y')}{2} \sigma_{i,j}^z \sigma_{i,j+1}^z}\\
    U_{h,i,j} &= e^{-\mathrm{i} h \sigma_{i,j}^x}\\
\end{split}
\end{equation}
can each be viewed as a quantum gate operation. In particular, the single-qubit rotation $U_{h,i,j}$ can be natively implemented with high fidelity in a typical superconductor circuit platform \cite{GoogleSupremacy}. The remaining two-qubit gates can be realized through a combination of single qubit rotations and iSWAP gates (iSWAP$=e^{\mathrm{i} \frac{\pi}{4} (\sigma^x \otimes \sigma^x +\sigma^y \otimes \sigma^y)}$), both of which are native within the platform of Ref.~\cite{GoogleSupremacy}. Indeed, observe that 
\begin{equation}
    e^{\mathrm{i} \frac{\pi}{4} \mathcal{I} \otimes \sigma^x} iSWAP^{(-1)} e^{\mathrm{i} \theta \sigma^x \otimes \mathcal{I}} iSWAP e^{-\mathrm{i} \frac{\pi}{4} \mathcal{I} \otimes \sigma^x} = e^{\mathrm{i} \theta \sigma^z \otimes \sigma^z }, 
\end{equation}
thereby realizing the desired $ZZ$ unitary appearing both in $U_{x,i,j}$ and $U_{y,i,j}$.

Having established a means to realize our system with superconducting qubits, we now discuss how its topological signature, i.e., the presence of ZMs and PMs, can be probed in the same setup. To this end, we first note the following two observations. First, both ZMs and PMs are localized near a system's corner, and their exact expressions can be obtained at special parameter values, i.e., Eq.~(\ref{idealmodes}). Second, note that a superposition of ZM and PM, e.g., $\gamma_0+\gamma_\pi$, exhibits $2T$-periodicity due to the $\pi$ phase difference between $\gamma_0$ and $\gamma_\pi$. Specifically,
\begin{equation}
    U^2 \left(\gamma_0+\gamma_\pi \right) \left(U^\dagger\right)^2 = U \left(\gamma_0-\gamma_\pi \right) U^\dagger = \gamma_0+\gamma_\pi \;.
\end{equation}
These observations can in turn be exploited as follows.

Consider again the ideal parameter values presented in Sec.~\ref{Sec: Model}, i.e., $h=\pi/4$, $J_x=\pi/2$, and $J_y'=0$, so that the ZMs and PMs are given by Eq.~(\ref{idealmodes}). It follows that the corner operators $\sigma_{1,1}^y$ and $\sigma_{1,1}^z$ precisely represent the superposition $\gamma_0+\gamma_\pi$ and $\gamma_0-\gamma_\pi$ respectively. From the argument above, $\sigma_{1,1}^y$ and $\sigma_{1,1}^z$ then turn into each other over the course of a period. Therefore, by preparing an initial state of the form $|00\cdots 0\rangle_z$ or $|00\cdots 0\rangle_y$, i.e., the $+1$ eigenstate of all $\sigma_{i,j}^z$'s or $\sigma_{i,j}^y$'s, the expectation value $\langle S_z \rangle (nT) \equiv \langle U^{n} \sigma_{1,1}^z U^{-n} \rangle$ alternates between $0$ and $+1$.

Let us now turn to the case of more general parameter values. Even if ZMs and PMs remain present, they no longer take the special forms of Eq.~(\ref{idealmodes}). Moreover, at finite system sizes, exact 0 and $\pi/T$ quasienergy excitations do not exist due to the unavoidable hybridization among ZMs and PMs at different corners. However, provided that the ZMs and PMs remain well-localized near a system's corner, they are expected to yield considerable overlap with $\sigma_{1,1}^z$. Consequently, the stroboscopic evolution of $\langle S_z \rangle (t=nT)$ will still display the transient oscillatory signature between values close to $0$ and $+1$ which can be observed in experiments. For a given set of parameter values at which the system still supports the ZMs and PMs, the lifetime of such oscillations generally improves with an increment in the system size. On the other hand, in the absence of ZMs and PMs, such oscillations either exhibit extremely short and size independent lifetimes, or may not even exist at all. A useful metric for quantifying such oscillations is the power spectrum $\langle \tilde{S}_z \rangle (\Omega) =|\frac{1}{L} \sum_{m=\ell}^L \langle S_z\rangle e^{-\mathrm{\frac{\ell \Omega T}{L}}}|$, obtained by Fourier transforming $\langle S_z \rangle$. In particular, the closer $\langle S_z \rangle$ from exhibiting $2T$-periodicity, the larger $\langle \tilde{S}_z \rangle (\pi/T)$ will be.

\begin{figure}
    \centering
    \begin{minipage}{0.23\linewidth}
		\begin{picture}(80,80)
			\put(-77,-20){\includegraphics[scale = 0.32]{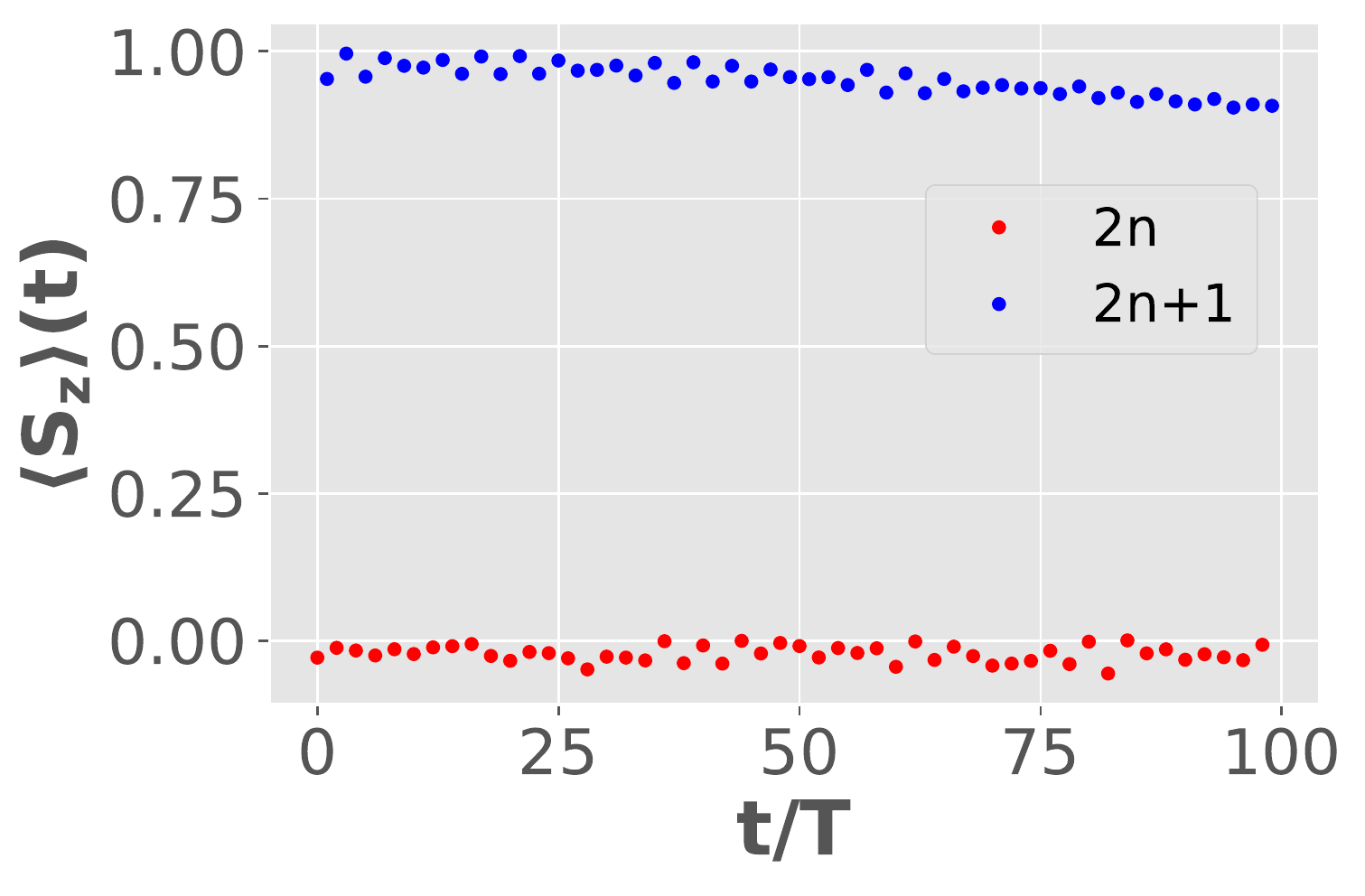}}		
			\put(-47,75){\textbf{\Large{(a)}}}	
		\end{picture}
	\end{minipage}
	~
	\begin{minipage}{0.23\linewidth}
		\begin{picture}(80,80)
			\put(-5,-20){\includegraphics[scale = 0.32]{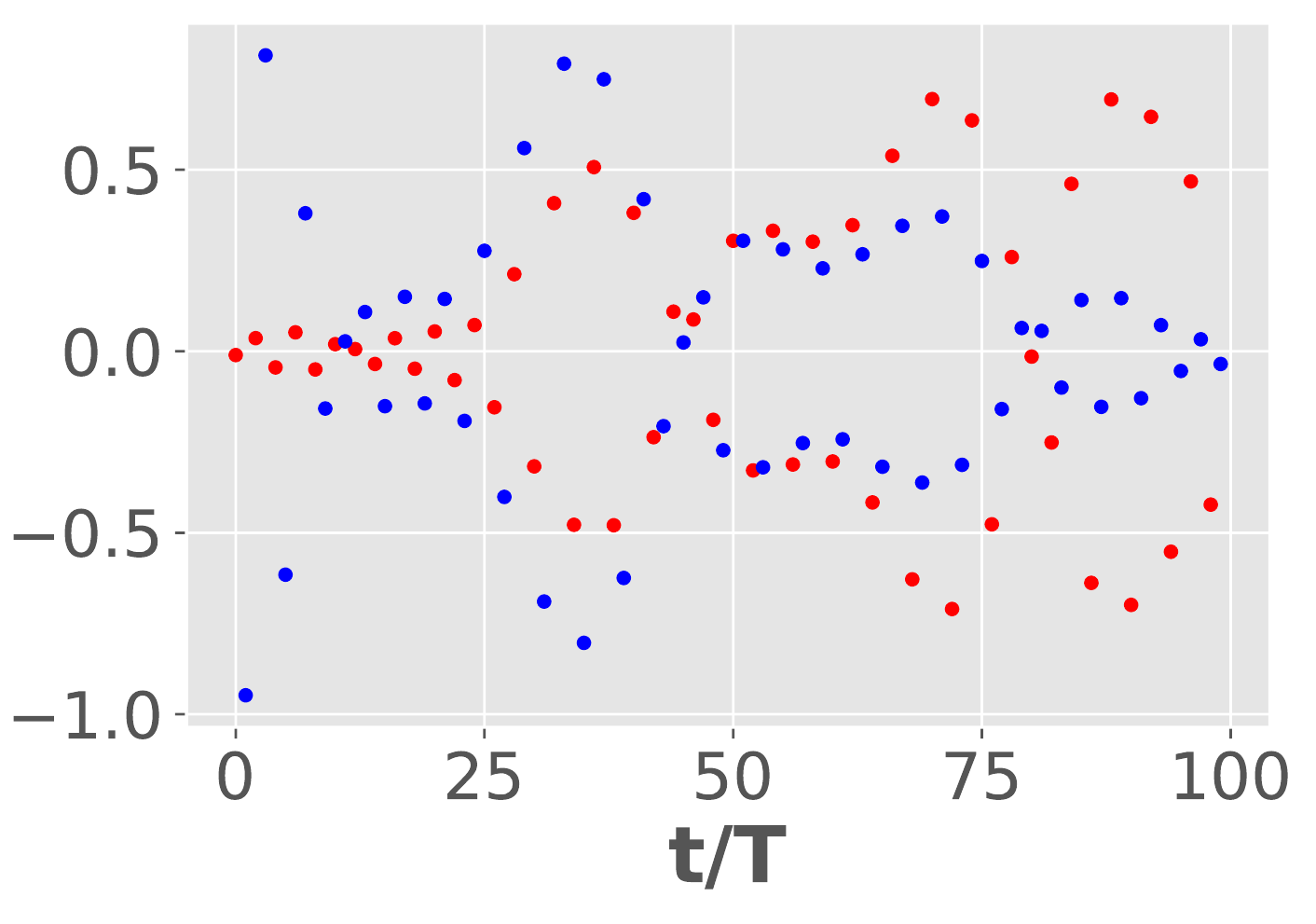}}	
			\put(15,75){\textbf{\Large{(b)}}}	
		\end{picture}
	\end{minipage}
	\\  
	\begin{minipage}{0.23\linewidth}
		\begin{picture}(80,80)
			\put(-74,-35){\includegraphics[scale = 0.32]{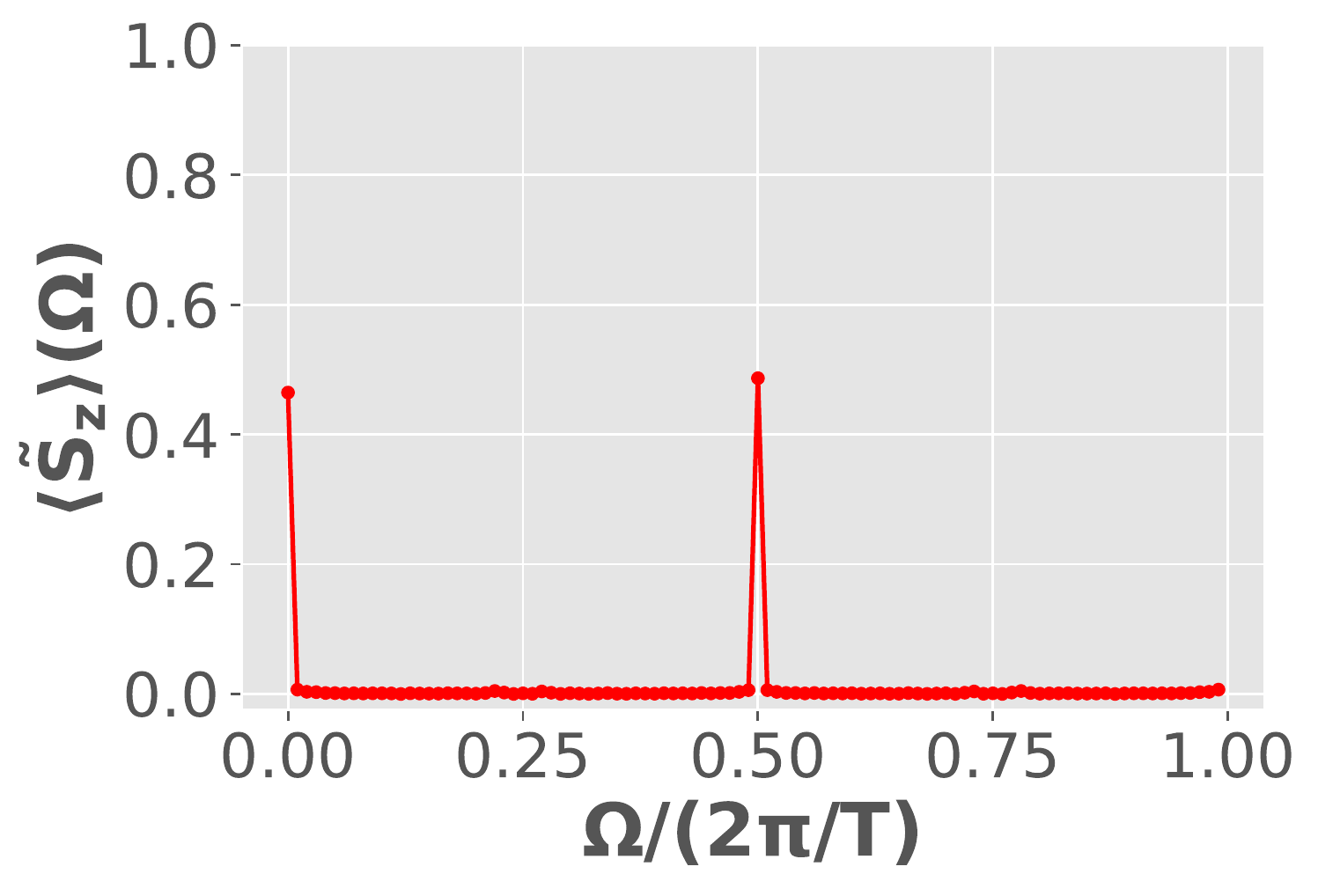}}		
			\put(-47,60){\textbf{\Large{(c)}}}	
		\end{picture}
	\end{minipage}
	~
	\begin{minipage}{0.23\linewidth}
		\begin{picture}(80,80)
			\put(0,-35){\includegraphics[scale = 0.32]{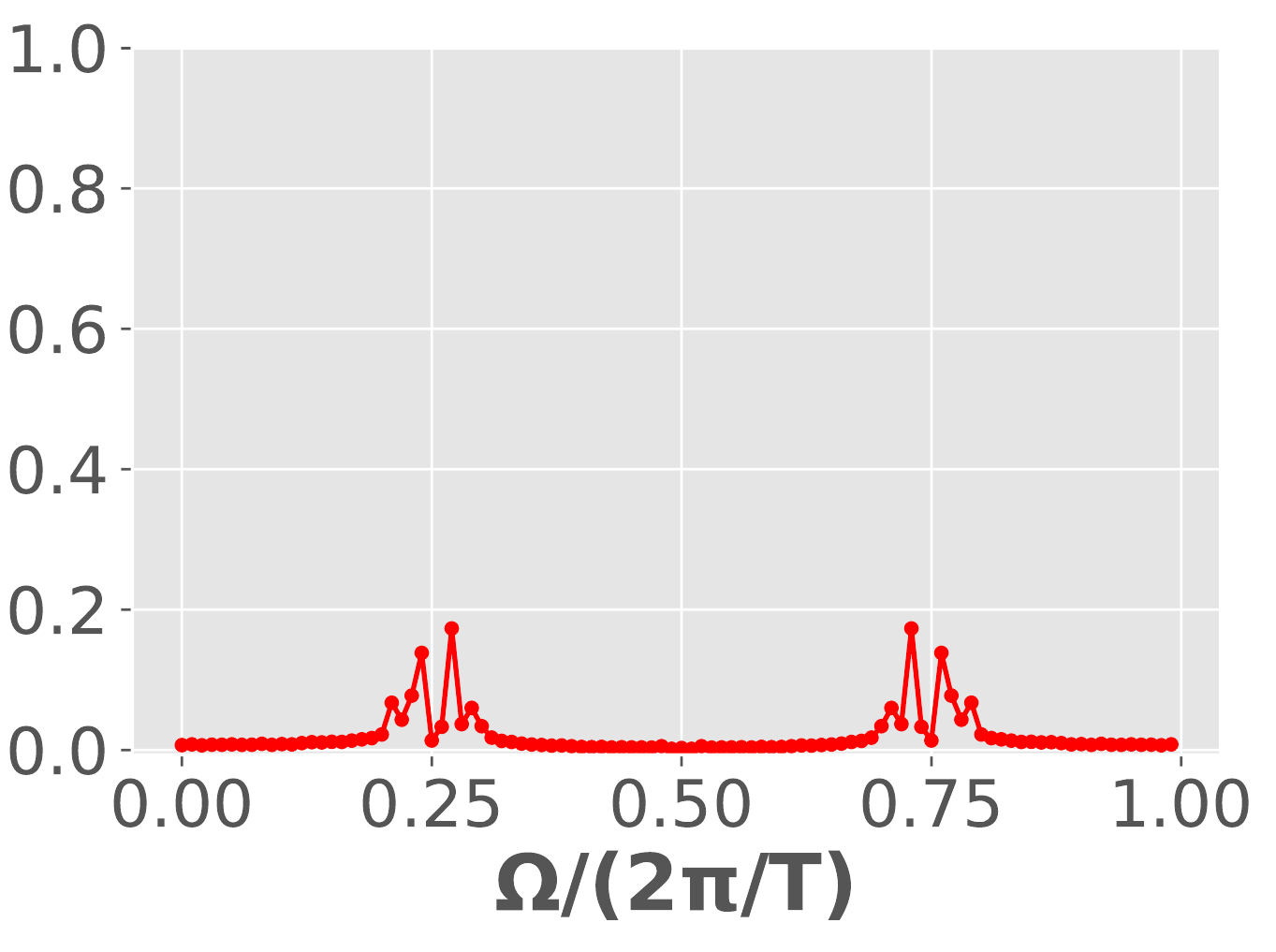}}
			\put(15,60){\textbf{\Large{(d)}}}	
		\end{picture}
	\end{minipage}
	\vspace{1.2cm}

    \caption{\textcolor{black}{(a,b) Stroboscopic evolution of a spin-1/2 particle residing at a corner of a lattice of size $4\times 4$, i.e., $\langle S_z \rangle (t)$, taking $|00\cdots 0\rangle_z$ as the initial state. Panels (c,d) depict the associated power spectrum. The parameter values are set to $2h=J_x=1.01\pi/2$, (a,c) $(J_y,J_y') = (1.1\pi/2,0.1\pi/2)$, (b,d) $(J_y,J_y') = (0.1\pi/2,1.1\pi/2)$. 
	}}
    \label{fig: Dynamics}
\end{figure}
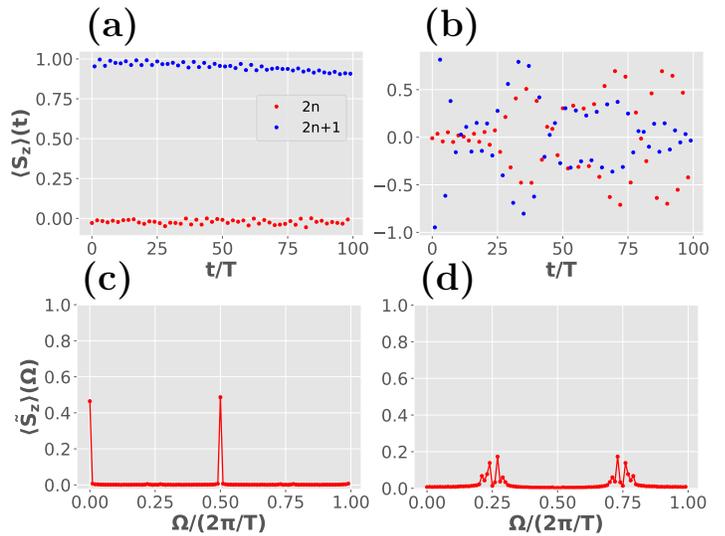

In Figs.~\ref{fig: Dynamics}(a,b), we numerically evaluate $\langle S_z \rangle(nT)$ at two representative sets of parameter values, respectively corresponding to the presence and absence of both ZMs and PMs. To simulate an actual superconducting circuit experiment, we evaluate such an expectation value by preparing a quantum circuit that represents $nT$ copies of Eq.~(\ref{Floquetcircuit}), running it via a Cirq package in Python \cite{cirq}, measuring the qubit $(1,1)$ in the computational basis, repeating the previous steps 10000 times, then finally averaging the measurement outcomes over these repetitions. As expected, a strong $2T$-period oscillation profile is only observed in panel (a), further supported by a very large $\Omega=\pi/T$ peak in the corresponding power spectrum plotted in panel (c). In contrast, such an oscillation profile is absent in panel (b) and its corresponding power spectrum in panel (d) instead shows two peaks centered around, but not at, $\Omega=\pi/T$. 

\begin{figure}
    \centering
    \includegraphics[scale=0.45]{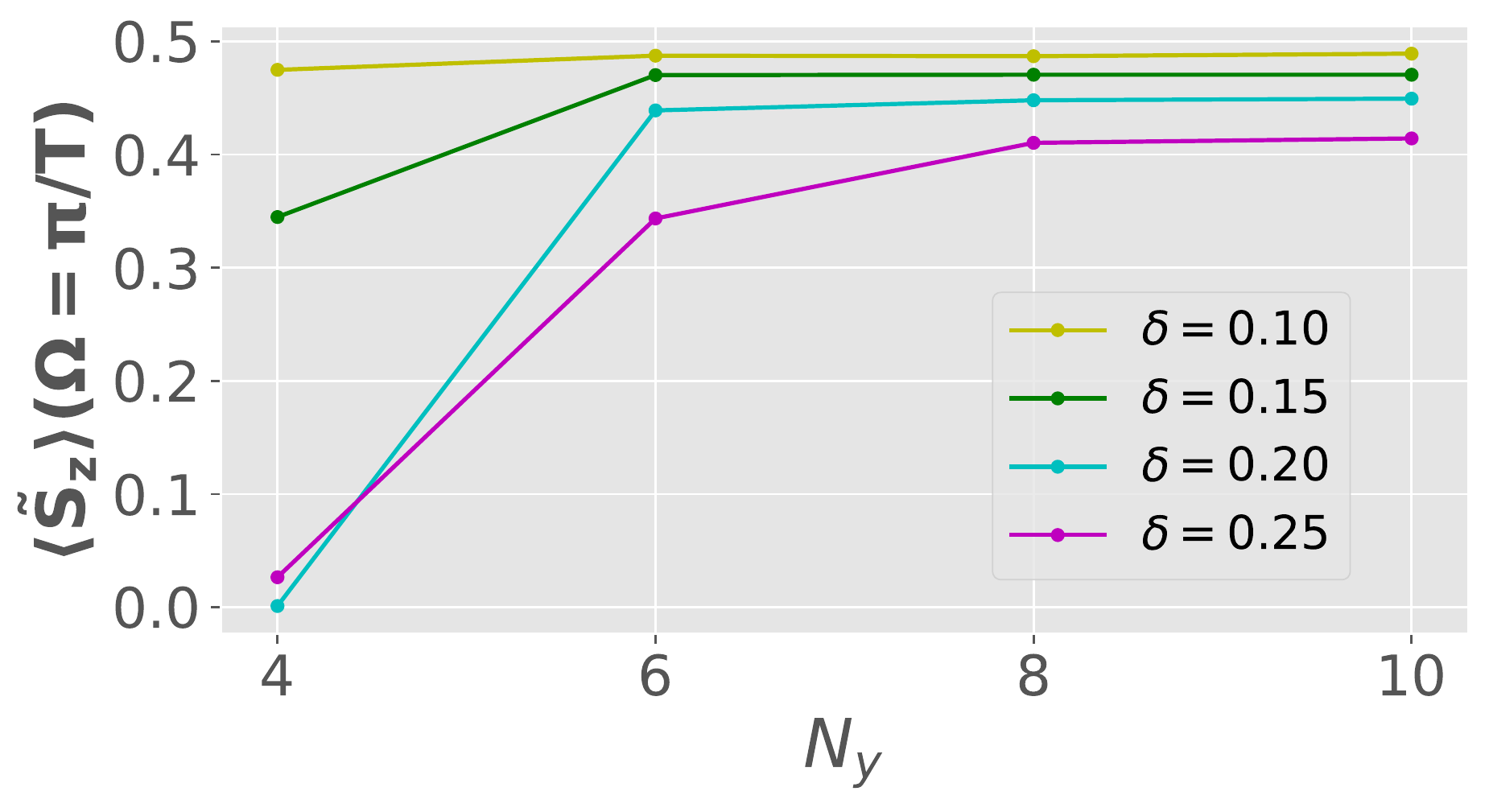}
    \caption{\textcolor{black}{Finite-Size Scaling. Here $2h=J_x=1.01\pi/2$. We have implemented various perturbations on $J_y$ and $J_y'$ away from their ideal values $(\pi/2,0)$. These are indicated by the parameter $\delta$, such that the values of $J_y,J_y'$ used are $(1+\delta)\pi/2$ and $\delta\pi/2$ respectively. The lattice is of size $2 \times N_y$.}}
    \label{finite size scaling}
\end{figure}

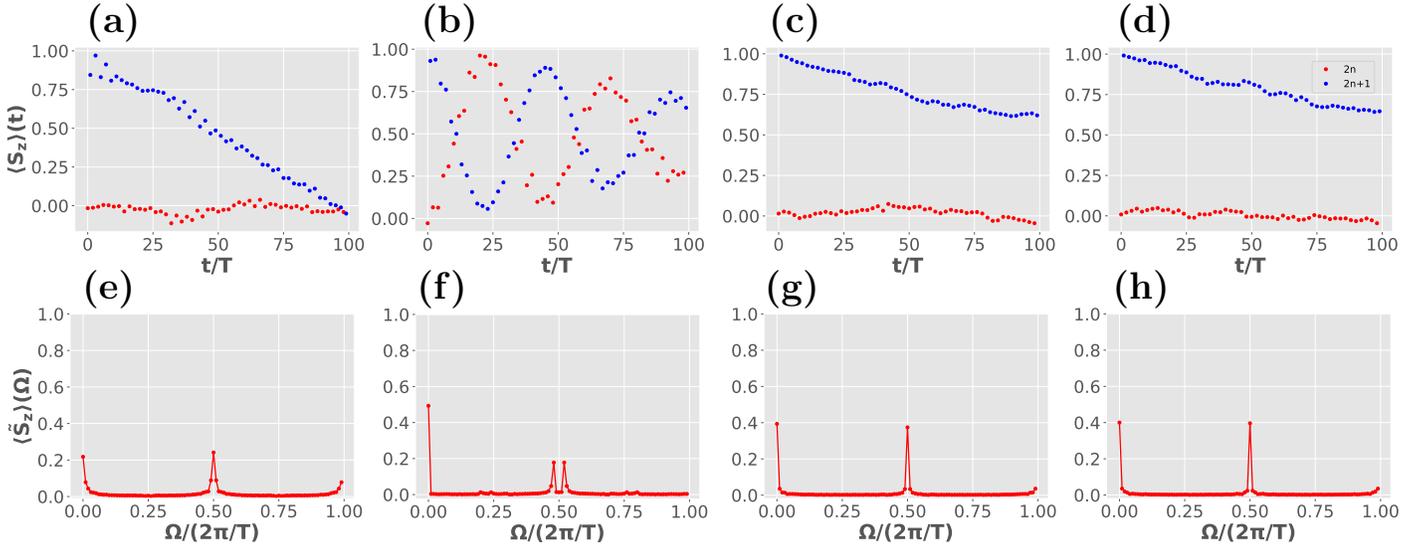
\begin{figure*}
	\centering
	\begin{minipage}{0.23\linewidth}
		\begin{picture}(80,80)
			\put(-42,-20){\includegraphics[scale = 0.32]{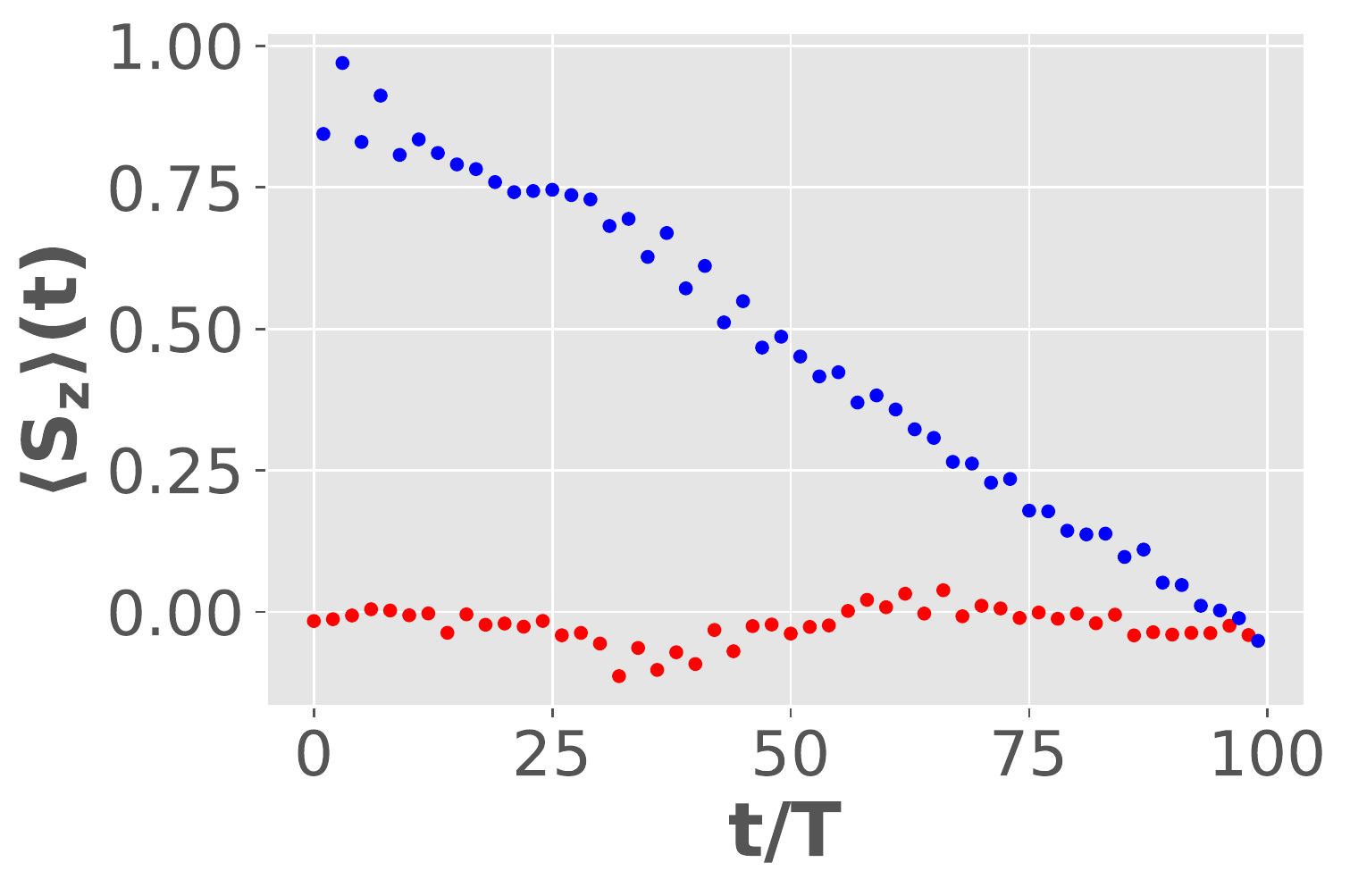}}		
			\put(-10,75){\textbf{\Large{(a)}}}	
		\end{picture}
	\end{minipage}
	~
	\begin{minipage}{0.23\linewidth}
		\begin{picture}(80,80)
			\put(-30,-20){\includegraphics[scale = 0.32]{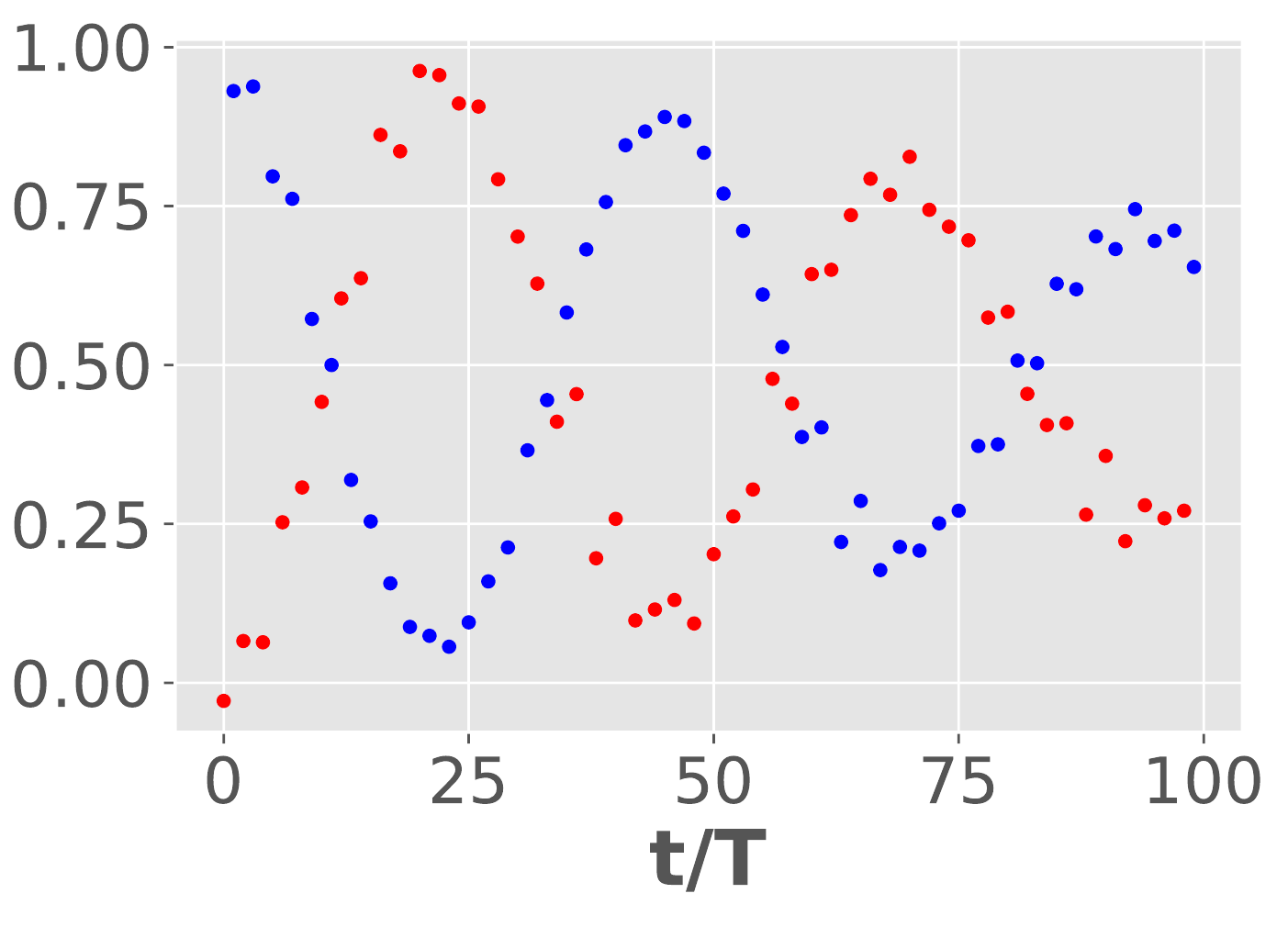}}	
			\put(-10,75){\textbf{\Large{(b)}}}	
		\end{picture}
	\end{minipage}
    ~
	\begin{minipage}{0.23\linewidth}
		\begin{picture}(80,80)
			\put(-25,-20){\includegraphics[scale = 0.32]{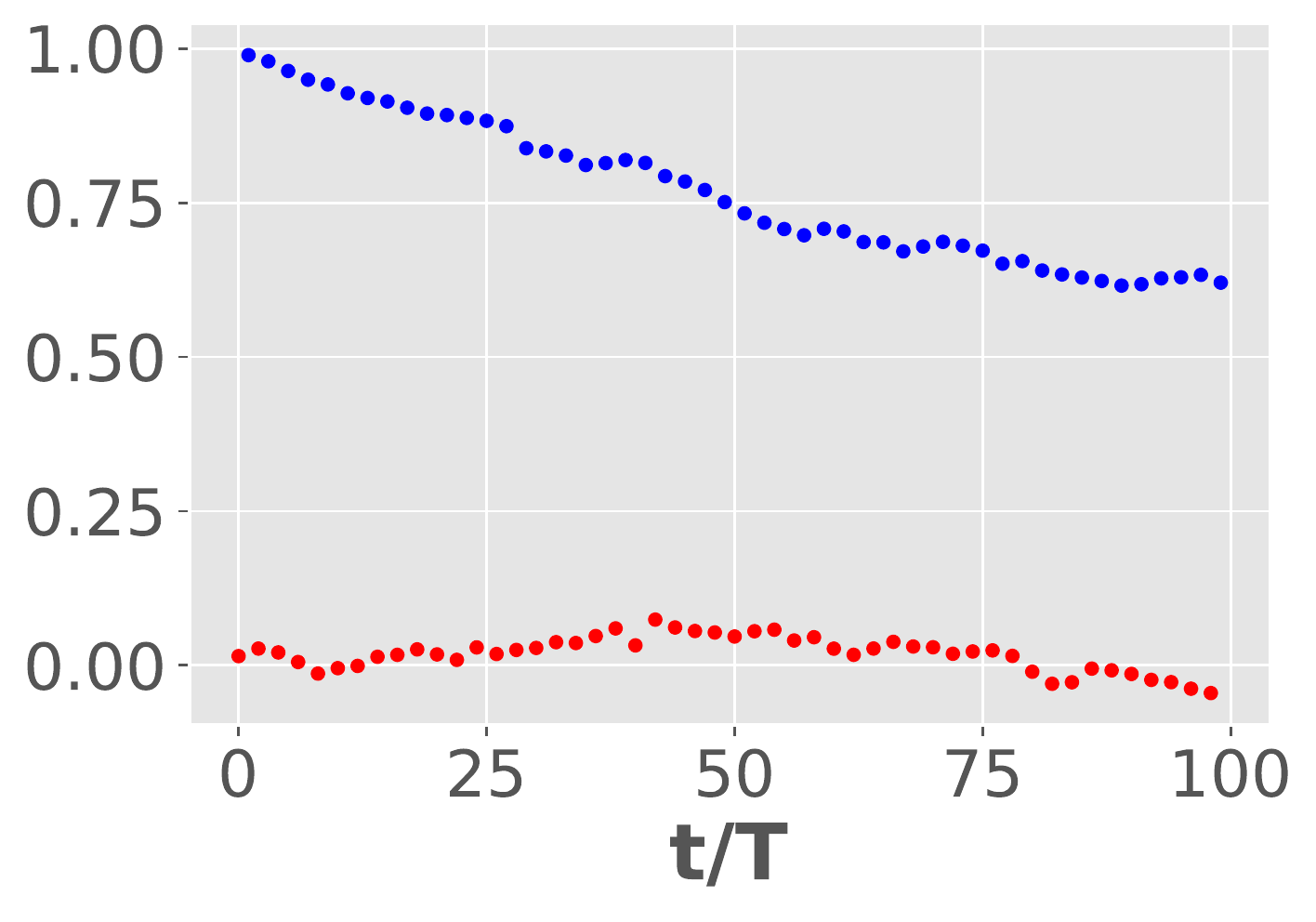}}		
			\put(-5,75){\textbf{\Large{(c)}}}	
		\end{picture}
	\end{minipage}
	~
	\begin{minipage}{0.23\linewidth}
		\begin{picture}(80,80)
			\put(-22,-20){\includegraphics[scale = 0.32]{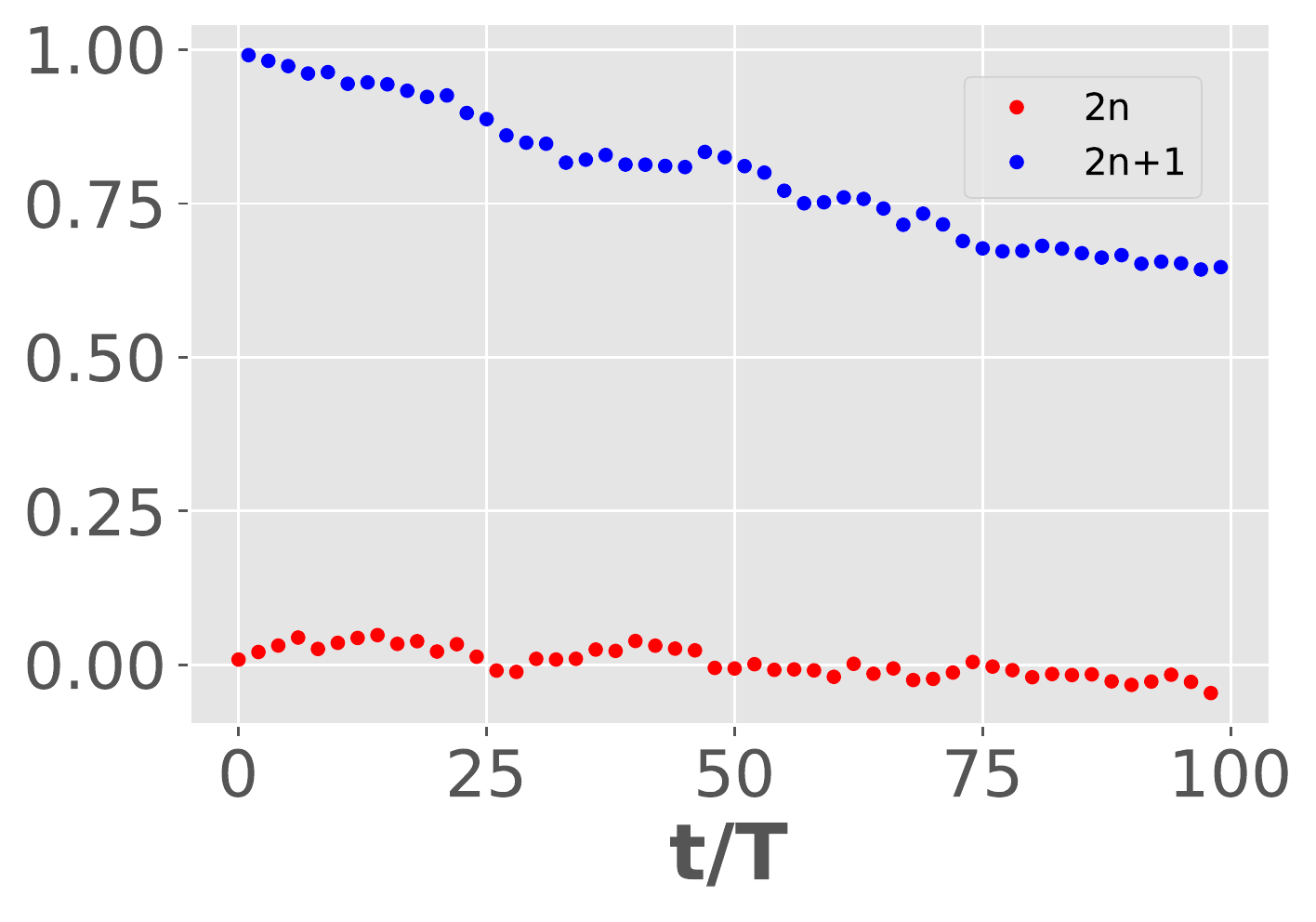}}		
			\put(0,75){\textbf{\Large{(d)}}}	
		\end{picture}
	\end{minipage}
	\\  
	\begin{minipage}{0.23\linewidth}
		\begin{picture}(80,80)
			\put(-40,-40){\includegraphics[scale = 0.32]{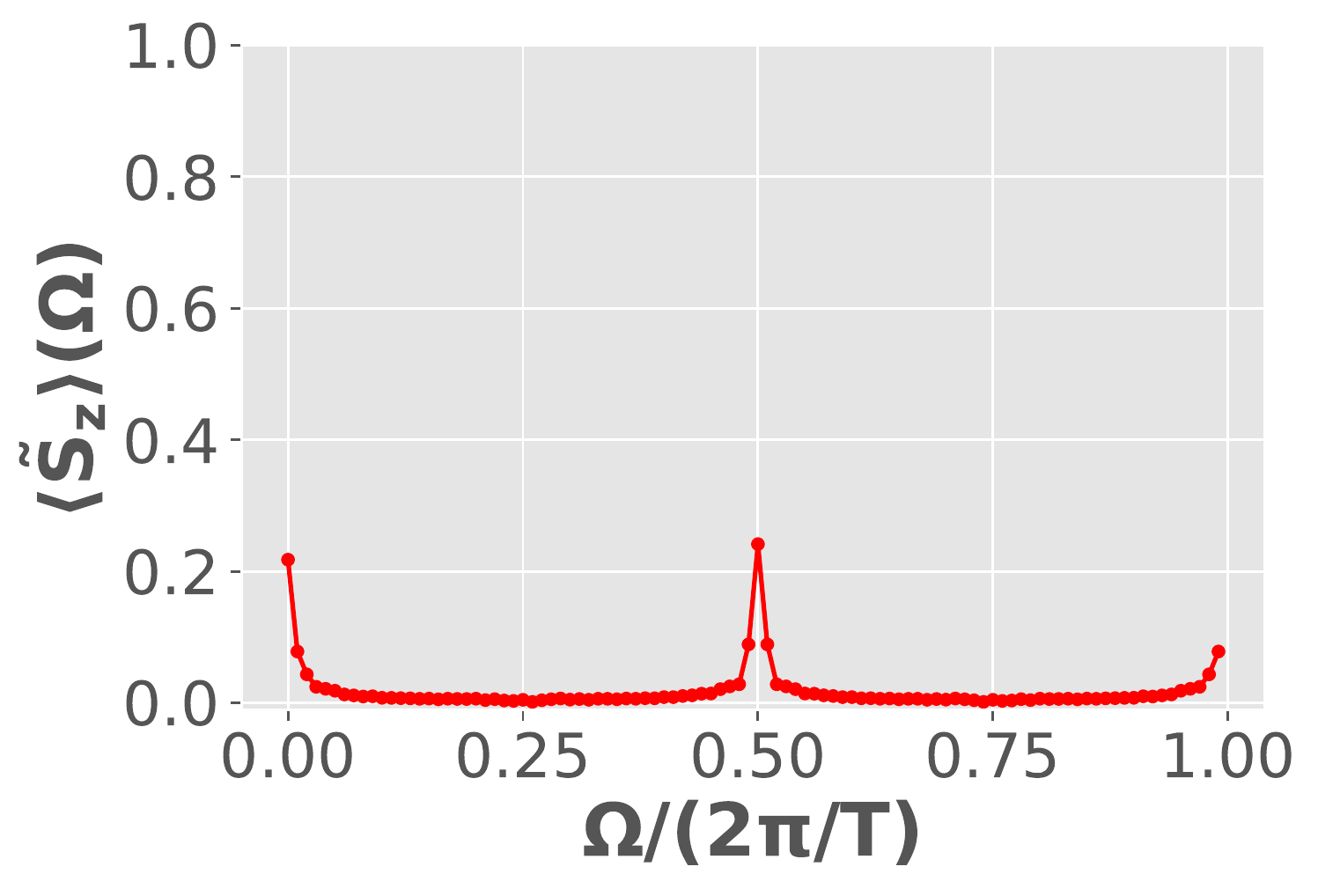}}		
			\put(-10,55){\textbf{\Large{(e)}}}	
		\end{picture}
	\end{minipage}
	~
	\begin{minipage}{0.23\linewidth}
		\begin{picture}(80,80)
			\put(-24,-40){\includegraphics[scale = 0.32]{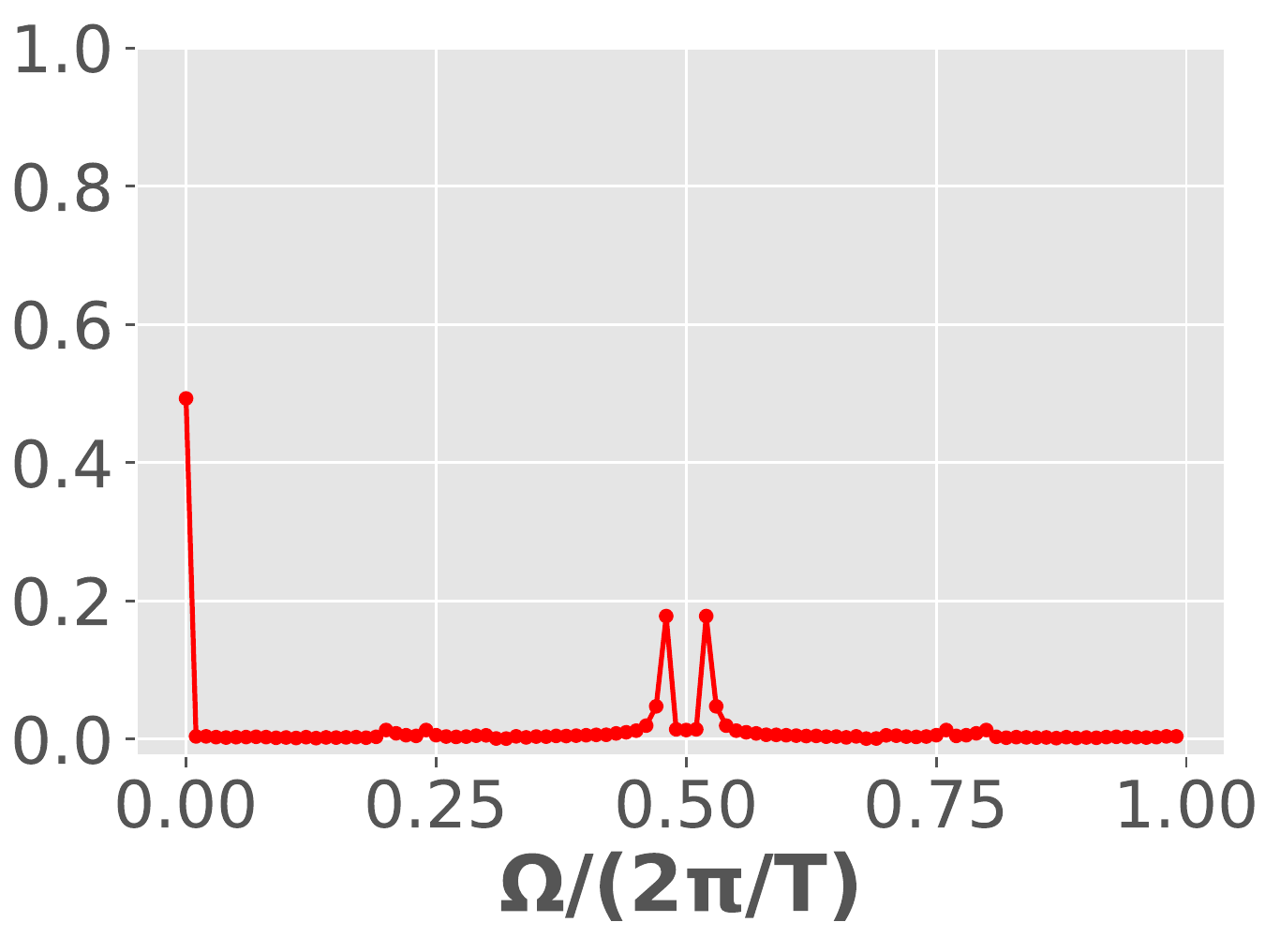}}
			\put(-10,55){\textbf{\Large{(f)}}}	
		\end{picture}
	\end{minipage}
    ~
	\begin{minipage}{0.23\linewidth}
		\begin{picture}(80,80)
			\put(-20,-40){\includegraphics[scale = 0.32]{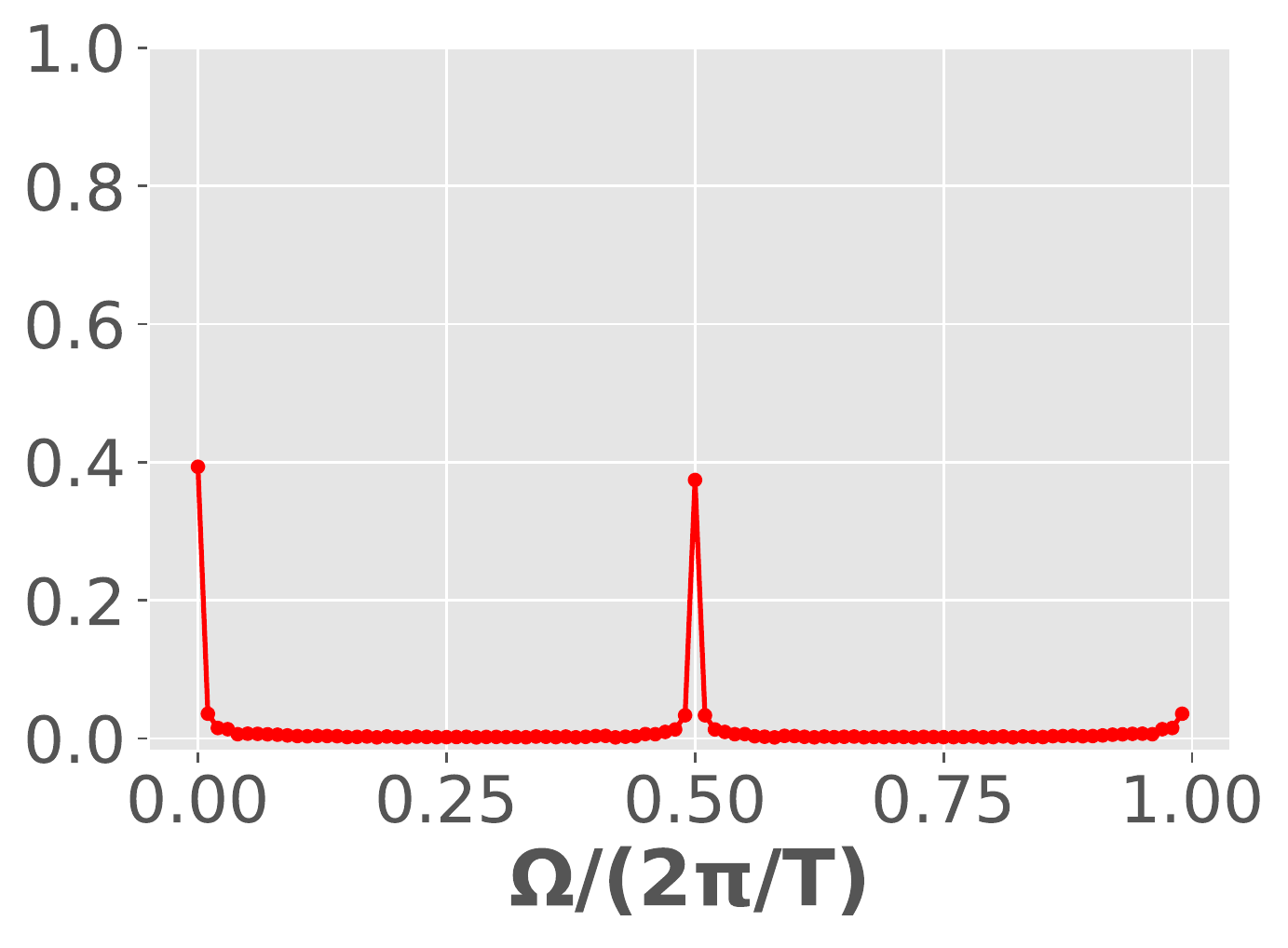}}
			\put(-5,55){\textbf{\Large{(g)}}}	
		\end{picture}
	\end{minipage}	
	~
	\begin{minipage}{0.23\linewidth}
		\begin{picture}(80,80)
			\put(-17,-40){\includegraphics[scale = 0.32]{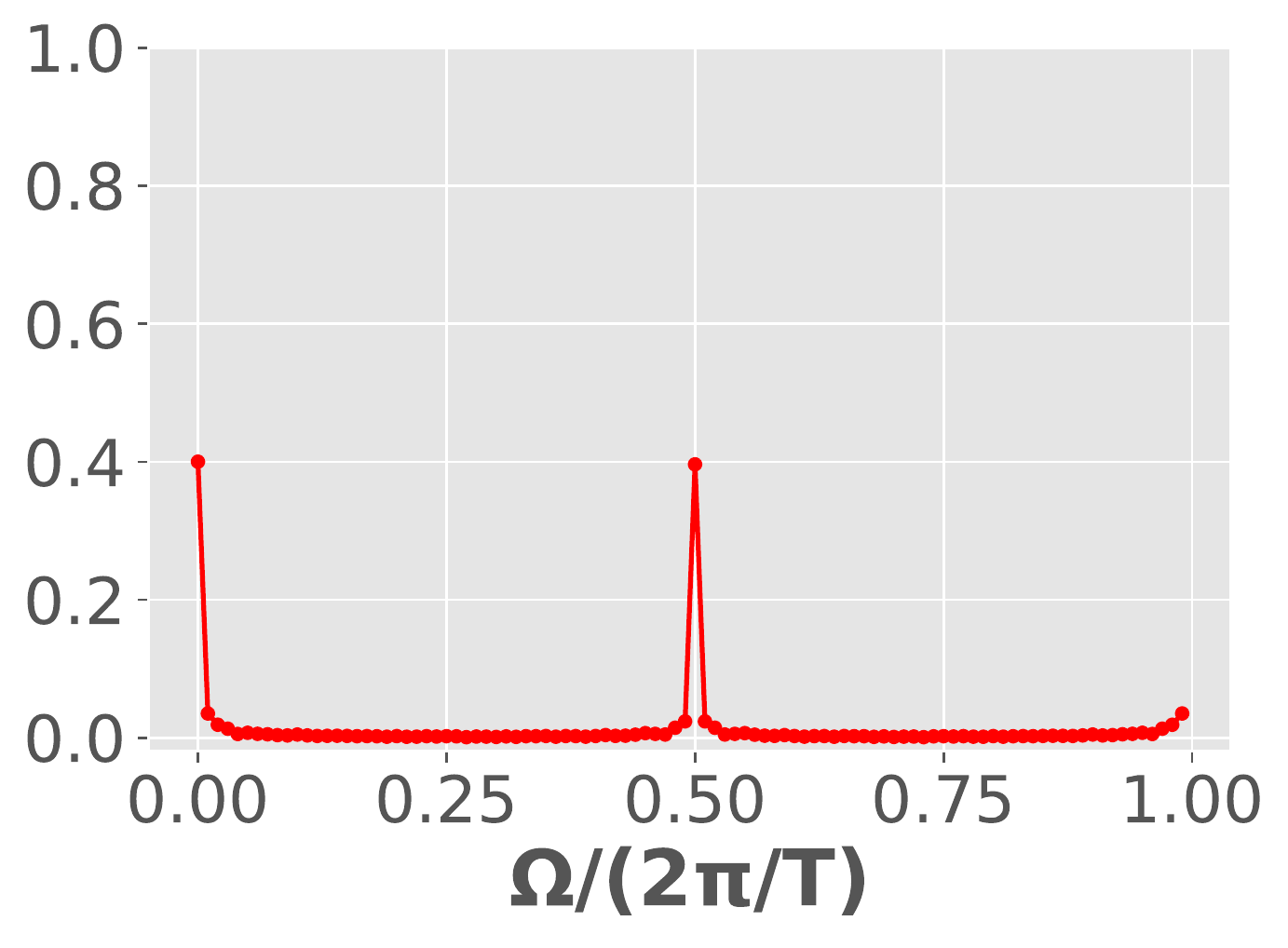}}		
			\put(0,55){\textbf{\Large{(h)}}}	
		\end{picture}
	\end{minipage}
	\vspace{1.5cm}
	
	\caption{\textcolor{black}{
	The role of imperfections.
	Perturbations: (a,e) $(J_y,J_y') = (1.1\pi/2,0.1\pi/2)$ with symmetry-preserving perturbation $\sum_{i=1}^{N_x-1} \sum_{j=1}^{N_y} \; s_p \Z_{ij}\Z_{i+1,j} + \sum_{i=1}^{N_x} \sum_{j=1}^{N_y-1} \; s_p \Z_{ij}\Z_{i,j+1}$, $s_p = 0.1$, added to the Hamiltonian, and (b,f) $(J_y,J_y') = (1.1\pi/2,0.1\pi/2)$ with symmetry-breaking perturbation $\sum_{i,j} \; s_b \Y_{ij}$, $s_b = 0.1$, added to the Hamiltonian. Here the lattice is of size $4 \times 4$. Disorder: (c,d) Stroboscopic evolution of $\langle S_z \rangle (t)$ and (g,h) its corresponding power spectrum in the presence of (c,g) spatial disorder of strength $0.05$, averaged over 25 realizations and (d,h) temporal disorder of strength $0.05$, also averaged over 25 realizations. In panels (c,d,g,h), the means of the system parameters are set to $\bar{h} = \pi/4$, $\bar{J}_x = \pi/2$, $\bar{J}_y = \pi/2$, and $\bar{J}_y' = 0$ (see also the caption in Fig. \ref{disorder scaling}). Here the lattice is of size $3 \times 4$.
	}}
    \label{fig: Dynamics with imperfections}
\end{figure*}

%\textcolor{red}{I modified the below paragraph a bit. It is not just $J_y'$ that is perturbed, $J_y$ is perturbed too. See caption in figures for details. I think you have to go through this again, since I'm not sure if I can simply replace all $J_y'$ by $\delta$.}

\RB{As the evaluation of time-evolution does not require exact diagonalization, the above method enables us to access a larger Hilbert space dimension as compared with the spectral functions studies presented in Sec.~\ref{numeric}. We may thus utilize such a dynamical approach to carry out a finite-size scaling analysis with respect to several perturbated $J_y$ and $J_y'$ values. To this end, we plot the subharmonic peak $\Omega=\pi/T$ of the power spectrum, which quantifies the tendency of $\langle S_z\rangle (t)$ to exhibit $2T$-periodicity, as a function of the relevant system size. In particular, for the same reason as that elucidated in Sec.~\ref{numeric}, we focus on varying only $N_y$ while fixing $N_x=2$. Our results are then summarized in Fig. \ref{finite size scaling}. There, we observe that for a sufficiently small perturbation strength (e.g. $\delta = 0.1$), $\langle \tilde{S}_z\rangle (\pi/T)$ is already near its maximum value of $0.5$ even at the smallest $N_y$ we considered, so that increasing the latter does not yield a significant effect. However, at larger perturbations (e.g. $\delta = 0.20, 0.25$), $\langle \tilde{S}_z\rangle (\pi/T)\approx 0$ at $N_y=4$, and it clearly increases with $N_y$. This is consistent with our expectation that the localization length of our ZMs and PMs in the $y$-direction depends on the perturbation $\delta$. In particular, the saturation in $\langle \tilde{S}_z\rangle (\pi/T)$ is achieved when $N_y$ is already larger than such a localization length, and the overlap between two opposite ZMs and PMs becomes negligible. That $\langle \tilde{S}_z\rangle (\pi/T)$ does not saturate to the maximum value of $0.5$ at large $\delta$ is attributed to the fact that our choice of $S_z$ is only approximately equal to the superposition of a ZM and a PM at such general parameter values.}

The above dynamical approach can further be exploited to investigate the effect of perturbations on the system's topology. In particular, we shall consider two types of perturbations, those that preserve and break the underlying $\mathbb{Z}_2$ symmetry of the system. In Figs.~\ref{fig: Dynamics with imperfections}(a,e), we find that the system demonstrates considerable robustness against symmetry-preserving perturbations, i.e., $2T$-oscillation pattern persists up to $t/T\approx 100$ under the chosen perturbation strength. On the other hand, the presence of symmetry-breaking perturbations proves to be detrimental; ZMs and PMs are no longer pinned at 0 and $\pi/T$ quasienergy excitations respectively, so their superpositions no longer yield a relative phase difference of $\pi$ over a period. This in turn results in a trivial oscillation profile observed in Figs.~\ref{fig: Dynamics with imperfections}(b,f), whose periodicity depends sensitively on some system parameters and is thus non-topological. These observations are consistent with the expectation that our system represents a symmetry-protected topological phase rather than a true topologically ordered phase.

\begin{figure}
    \centering
    \includegraphics[scale=0.45]{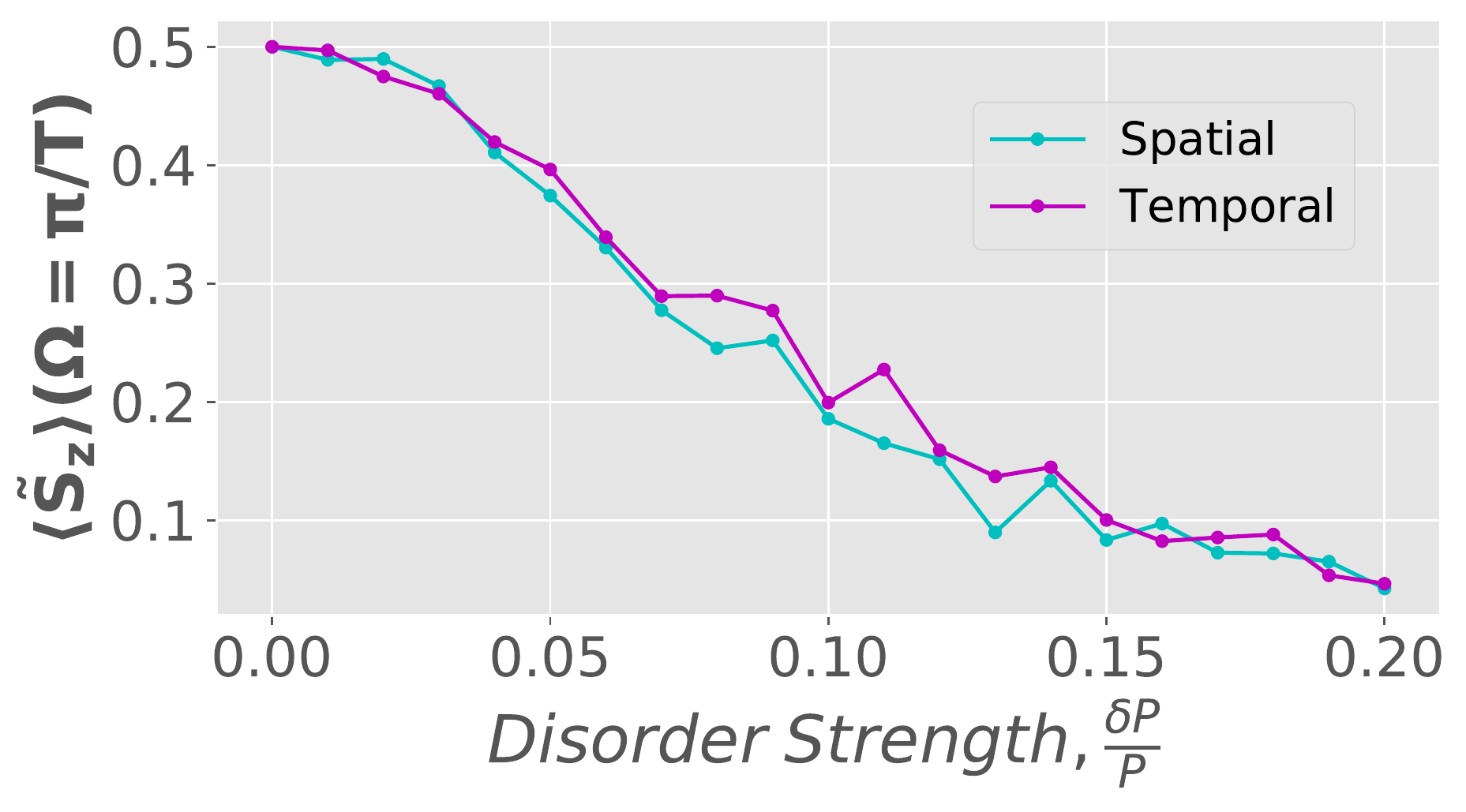}
    \caption{\textcolor{black}{Disorder Scaling. Here the lattice is of size $3 \times 4$, as in Figs. \ref{fig: Dynamics with imperfections}(c,d,g,h). Two types of disorder - spatial and temporal - implemented for all $h,J_x,J_y,J_y'$ are shown separately. Specifically, except for $J_y'$, each parameter is drawn from the uniform distribution as $P \in [\overline{P}-\delta P,\overline{P}+\delta P]$ under the same disorder strength $\delta P/P\equiv c$, where $(\overline{h},\overline{J}_x,\overline{J}_y)=(\pi/4,\pi/2,\pi/2)$. On the other hand, $J_y'$ is taken from the uniform distribution $[-c\pi/2+c\pi/2]$, which is of the same size as that of $J_y$.}} %The disorder strength for the parameter $P\in \left\lbrace h, J_x, J_y \right\rbrace$ is denoted by $\delta P/P$. This means $P \in [\overline{P}-\delta P,\overline{P}+\delta P]$. Here, $(\overline{h},\overline{J}_x,\overline{J}_y,\overline{J}_y')=(\pi/4,\pi/2,\pi/2,0)$. The parameter $J_y'$ is drawn randomly from the interval $[-\delta J_y/J_y,\delta J_y/J_y]$. Since this happens to $J_y'$, to rectify this we choose the size of its window to be the same as that of $J_y$. That is, supposing the disorder is $\delta P/P=c$, then since $J_y \in [\pi/2-c\pi/2,\pi/2+c\pi/2]$, we set $J_y' \in [-c\pi/2,+c\pi/2]$. These two windows have the same sizes, but are centered at different points.}}
    \label{disorder scaling}
\end{figure}

Finally, we note that in actual experiments, the perfect implementation of Eq.~(\ref{Floquetcircuit}) is generally impossible. In particular, any single- and two-qubit gate realizable with the superconducting circuit platform of Ref. \cite{GoogleSupremacy} generally has a high but less than unity fidelity, which in turn results in a disordered version of our system. To investigate how such imperfections potentially affect the experimental observation of ZMs and PMs in our system, we explicitly carry out another dynamical study of our system in the presence of spatial and temporal disorder. In the implementation of spatial disorder, all system parameters are made site-dependent, the values of which are drawn from a uniform distribution, i.e., $P_{i,j}\in [\bar{P}-\delta P, \bar{P}+\delta P]$ for $P=h,J_x,J_y,J_y'$. On the other hand, temporal disorder amounts to keeping all system parameters site-independent, but with values differing over different Floquet cycles. In this case, the time evolution operator over $n$ periods takes the form $\prod_{m=1}^n U_m$, where each $U_m$ satisfies Eq.~(\ref{Floquetcircuit}) with all system parameters drawn from a uniform distribution, $P_m\in [\bar{P}-\delta P, \bar{P}+\delta P]$ for $P=h,J_x,J_y,J_y'$. In both cases of spatial and temporal disorder, $\delta P/P$ is referred to as the disorder strength. Our results, as summarized in Figs.~\ref{fig: Dynamics with imperfections}(c,d,g,h) and Fig. \ref{disorder scaling}, demonstrate that ZMs and PMs are in fact robust against both disorder effects. Consequently, the dynamical approach proposed in this section indeed serves as a feasible means to detect our system's topology.

\section{Concluding Remarks}
\label{Conclusion}

We have constructed a relatively simple interacting and periodically driven 2D spin-1/2 system which supports an anomalous symmetry-protected second-order topological phase. The latter leads to the simultaneous presence of corner localized ZMs and PMs that spontaneously break the system's underlying $\mathbb{Z}_2$ symmetry. Despite being a genuinely interacting system, analytical topological characterizations of these ZMs and PMs can be made in some special cases. We further numerically computed the appropriate spectral functions to determine the presence of corner ZMs and PMs in the regime inaccessible by analytical treatment.

We have discussed the potential implementation of our system in existing superconducting circuit platforms. In addition, we also proposed a means to probe the signature of ZMs and PMs in such experiments by inspecting the dynamics of a qubit residing at a corner. Finally, we demonstrated the robustness of these ZMs and PMs against symmetry-preserving perturbations and disorder effects.

%\textcolor{red}{
%We have constructed a system subject to both periodic driving and dimerization, and demonstrated, via both analytical and numerical approaches the existence of second-order symmetry protected topological phases in the system, under certain conditions, which turned out to be a inequality relating the system parameters. This inequality was first obtained by analyzing eigenvalue equations in operator space. This result is further supported by analyzing relevant topological invariants. On the numerical side, we have evaluated the spectral function phase diagrams, which are to be compared with the analytical phase diagram described by the inequality. Of more practical relevance to experimentalists, the simulation of the stroboscopic evolution of a single spin residing at the corner was also carried out. All four approaches complement each other well, and led us to the present conclusion.
%}

There are several interesting avenues for future work. First, incorporating longer-range interactions in the present system is expected to yield richer physics. Indeed, many interesting phenomena, such as those expected from black-hole physics \cite{Rosenhaus}, disorder-free discrete time crystals \cite{Russomanno_etal}, and low-density parity-check quantum error correction codes \cite{BE}, have been proposed to arise in strictly long-range systems. Second, replacing the spin-1/2 Pauli matrices in the present system with their higher-spin generalizations \cite{Fendley1,ZhouRadWu} may lead to a family of exotic Floquet $\mathbb{Z}_n$ symmetry-protected topological phases. In particular, such systems are expected to support the more elusive $2\pi/n$ modes \cite{Sreejith_etal} at their corners, generalizing the corner PMs achievable in the present system. Finally, the coexistence of ZMs and PMs in the present system may be exploited for quantum computing applications, e.g., in the spirit of \cite{RadityaGong2, RadityaGong3}.

\begin{acknowledgments}
	R.W.B. is supported by the Australian Research Council Centre of Excellence for Engineered Quantum Systems (EQUS, CE170100009). K.K. and K.L.C. are supported by the Ministry of Education and the National Research Foundation under the Center for Quantum Technologies (CQT).
	
	K.K. and R.W.B. contributed equally to this work.
\end{acknowledgments}

%\clearpage
%\newpage

\appendix
\section{Analytical Derivation of the Phase Diagram}
\label{app1}

Here, we elaborate the detailed derivation of the phase diagram in Fig.~\ref{fig: Main Phase Diagram} of the main text. To this end, we remind the readers that we fix $(h,J_x)=(\pi/4,\pi/2)$ throughout this section.

Let us first set up some notation and terminology. Let $f$ be a function with domain $A$ and codomain $B$. We shall denote the space of all such functions as $\text{Func}(A,B)$. If $f: A \longrightarrow B$ is a linear map on vector spaces, we shall also write $\text{Func}(A,B)$ as $\text{Hom}(A,B)$. Next denote our spin lattice as $\mathbb{L} = \{(i,j) \;|\; i \in \{1 \dots N_x\}, j \in \{1 \dots N_y\}\}$, and its Hilbert Space as $\mathbb{H}_{\mathbb{L}}$. At each site, the operator space is spanned by the four standard Pauli Matrices $\sigma^{0/1/2/3}$, where as usual, the convention is $\sigma^{1/2/3} \longleftrightarrow \sigma^{x/y/z}$ and $\sigma^0=\mathcal{I}$. The operator space of the lattice $\mathbb{O}_\mathbb{L} = \text{Hom}(\mathbb{H}_{\mathbb{L}},\mathbb{H}_{\mathbb{L}})$ is then spanned by the (generalized) Pauli Basis
\[
    \left\{\prod_{(i,j) \in \mathbb{L}} \sigma_{i,j}^{p(i,j)} \;|\; p \in \text{Func}(\mathbb{L},\{0,1,2,3\}) \right\}
\]
where the map $p: \mathbb{L} \longrightarrow \{0,1,2,3\}$ assigns to each lattice site a Pauli Matrix.

A general element of $\mathbb{O}_{\mathbb{L}}$, i.e. a linear operator $O$ acting on $\mathbb{L}$ can be written as
\[
    O \;= \sum_{p \in \text{Func}(\mathbb{L},\{0,1,2,3\})} c_p \prod_{(i,j) \in \mathbb{L}} \sigma_{i,j}^{p(i,j)}
\]
In particular, if $O$ is Hermitian, $c_p$ are all real. An example of such operators is our system Hamiltonian. {\it A priori}, all elements of $\mathbb{O}_{\mathbb{L}}$ that do not commute with the system's $\mathbb{Z}_2$ symmetry $S$ are possible candidates for $0/\pi$-modes. However, for such modes to be topological in nature, they must be localized near a corner, i.e., $c_p$ converges exponentially to zero at sites away from the corner. In the following, we specifically look for such corner-localized operators, that further satisfy $U \mathbb{O} U^\dagger = \mathbb{O}$ or $U \mathbb{O} U^\dagger = -\mathbb{O}$ for ZMs or PMs respectively. To begin, we define the superoperator $F \in \text{Hom}(\mathbb{O}_{\mathbb{L}},\mathbb{O}_{\mathbb{L}})$ by
\[
    F: O \longrightarrow UOU^{\dag},
\]
which represents the action of conjugating $O$ by the Floquet operator. By diagonalizing $F$, the ZMs and PMs thus correspond precisely to the eigenvectors of $F$ with corresponding eigenvalues of $+1$ and $-1$ respectively. Finally, we also define the shorthand notations $c_y \equiv \cos(2J_y)$, $s_y \equiv \sin(2J_y)$, $c_y' \equiv \cos(2J_y')$, and $s_y' \equiv \sin(2J_y')$ throughout this section. 

We will now present our construction of a ZM and PM localized near $(1,1)$. The same steps can be repeated straightforwardly to construct ZMs and PMs near the other three corners. Our strategy is to first evaluate how $F$ acts on a corner Pauli operator $P_{11}$. Under the special parameter values of $h$ and $J_x$ under consideration, $F(P_{11})$ only involves a linear combination of Pauli matrices residing at sites $(1,1)$ and $(1,2)$. We then further evaluate how $F$ acts on each of these Pauli matrices, the results of which can be written down as linear combinations of Pauli matrices residing at the sites $(1,1)$, $(1,2)$, and $(1,3)$. By repeating this procedure, we identify a set of Pauli matrices residing at sites $(1,1)$, $(1,2)$, $(1,3)$, $...$ , $(1,N_y)$ that are closed under $F$. Restricting $F$ to the subspace spanned by the specific set of Pauli matrices we have identified then yields a $2L_y\times 2L_y$ matrix that is possible to diagonalize. Finally, the $\pm 1$ eigenvalues and eigenvectors of such a matrix provide information on the existence of ZMs and PMs.  

%A priori, we expect these terms to run through (linear combinations of) the entire Pauli Basis, i.e the number of terms involved is on the order of $O(2^n)$. But it turns out that only a small subset of the Pauli Basis, on the order of $O(n)$ is involved. That means that if we write down the matrix representation of $F$ w.r.t. the Pauli Basis, an $n \times n$ matrix suffices, instead of the usual $2^n \times 2^n$.

%Furthermore, this matrix representation has a nice pattern, and enables one to solve for these modes analytically, which leads to the simple result in \ref{MainResult}.

We will now mathematically elaborate the above argument. First, note that $I_{11},\X_{11}$ can be discarded immediately from the initial choice of $P_{11}$, since $[S,I]=[S,\X]=0$. On the other hand, applying $F$ to $\Y_{11}$ and $\Z_{11}$ gives
\begin{align*}
    F(\Y_{11}) &= \Z_{11}\\
    F(\Z_{11}) &= c_y'\Y_{11} - s_y'\X_{11}\Z_{12}. 
\end{align*}
According to our outlined strategy, we then apply $F$ to $\X_{11}\Z_{12}$ to obtain $F(\X_{11}\Z_{12}) = -c_y\X_{11}\Y_{12}+s_y\X_{11}\X_{12}\Z_{13}$. We further evaluate $F(\X_{11}\Y_{12})$ and $F(\X_{11}\X_{12}\Z_{13})$ and so forth. We see that there is a recurring pattern:
\begin{enumerate}
    \item \underline{$n$ odd ($n+1$ even)}
    \begin{align*}
    F\left[\left(\prod_{i=1}^n \X_{1i}\right) \Y_{1,n+1}\right] = &-c_y'\left(\prod_{i=1}^n \X_{1i}\right)\Z_{1,n+1}\\ &-s_y'\left(\prod_{i=1}^{n-1} \X_{1i}\right)\Y_{1n}\\
    F\left[\left(\prod_{i=1}^n \X_{1i}\right) \Z_{1,n+1}\right] = &-c_y\left(\prod_{i=1}^n \X_{1i}\right)\Y_{1,n+1}\\ 
    &+ s_y\left(\prod_{i=1}^n \X_{1i}\right)\X_{1,n+1}\Z_{1,n+2}
    \end{align*}
    \item \underline{$n$ even ($n+1$ odd)}
    \begin{align*}
    F\left[\left(\prod_{i=1}^n \X_{1i}\right) \Y_{1,n+1}\right] =\;\; &c_y\left(\prod_{i=1}^n \X_{1i}\right)\Z_{1,n+1}\\ &+s_y\left(\prod_{i=1}^{n-1} \X_{1i}\right)\Y_{1n}\\
    F\left[\left(\prod_{i=1}^n \X_{1i}\right) \Z_{1,n+1}\right] =\;\; &c_y'\left(\prod_{i=1}^n \X_{1i}\right)\Y_{1,n+1}\\ 
    &- s_y'\left(\prod_{i=1}^n \X_{1i}\right)\X_{1,n+1}\Z_{1,n+2}
    \end{align*}
\end{enumerate}
Note that the Pauli Operators involved are propagated down vertically, and take on only the form $(\prod \sigma^x)(\sigma^{y/z})$. Finally, we also have $F((\prod_{i=1}^{N_y-1} \X_i)\Z_{N_y}) = -(\prod_{i=1}^{N_y-1} \X_i)\Y_{N_y}$.

We identify that the set
\begin{align*}
    \mathbb{B} =  \{&\Y_{11},\Z_{11},\\
                    &\X_{11}\Y_{12},\X_{11}\Z_{12},\\
                    &\X_{11}\X_{12}\Y_{13},\X_{11}\X_{12}\Z_{13},\\
                    &\vdots\\
                    &\X_{11}...\X_{1,N_y-1}\Y_{1N_y},\X_{11}...\X_{1,N_y-1}\Z_{1N_y}\}
\end{align*}
is closed under $F$. The restriction of $F$ to the subspace spanned by $\mathbb{B}$ can then be written (in matrix representation) as (taking $N_y=8$, say)
\begin{widetext}
\begin{equation}\label{Fmatrix}
    \mathbf{F} = 
    \begin{bmatrix}
    & \cdot & c_y' & -s_y' & \cdot & \cdot & \cdot & \cdot & \cdot & \cdot & \cdot & \cdot & \cdot & \cdot & \cdot & \cdot & \cdot &\\
    & 1 & \cdot & \cdot & \cdot & \cdot & \cdot & \cdot & \cdot & \cdot & \cdot & \cdot & \cdot & \cdot & \cdot & \cdot & \cdot &\\
    & \cdot & \cdot & \cdot & \textcolor{blue}{\mathbf{-c_y}} & \textcolor{blue}{\mathbf{s_y}} & \cdot & \cdot & \cdot & \cdot & \cdot & \cdot & \cdot & \cdot & \cdot & \cdot & \cdot &\\
    & \cdot & \textcolor{blue}{\mathbf{-s_y'}} & \textcolor{blue}{\mathbf{-c_y'}} & \cdot & \cdot & \cdot & \cdot & \cdot & \cdot & \cdot & \cdot & \cdot & \cdot & \cdot & \cdot & \cdot &\\
    & \cdot & \cdot & \cdot & \cdot & \cdot & \textcolor{red}{\mathbf{c_y'}} & \textcolor{red}{\mathbf{-s_y'}} & \cdot & \cdot & \cdot & \cdot & \cdot & \cdot & \cdot & \cdot & \cdot &\\
    & \cdot & \cdot & \cdot & \textcolor{red}{\mathbf{s_y}} & \textcolor{red}{\mathbf{c_y}} & \cdot & \cdot & \cdot & \cdot & \cdot & \cdot & \cdot & \cdot & \cdot & \cdot & \cdot &\\
    & \cdot & \cdot & \cdot & \cdot & \cdot & \cdot & \cdot & \textcolor{blue}{\mathbf{-c_y}} & \textcolor{blue}{\mathbf{s_y}} & \cdot & \cdot & \cdot & \cdot & \cdot & \cdot & \cdot &\\
    & \cdot & \cdot & \cdot & \cdot & \cdot & \textcolor{blue}{\mathbf{-s_y'}} & \textcolor{blue}{\mathbf{-c_y'}} & \cdot & \cdot & \cdot & \cdot & \cdot & \cdot & \cdot & \cdot & \cdot &\\
    & \cdot & \cdot & \cdot & \cdot & \cdot & \cdot & \cdot & \cdot & \cdot & \textcolor{red}{\mathbf{c_y'}} & \textcolor{red}{\mathbf{-s_y'}} & \cdot & \cdot & \cdot & \cdot & \cdot &\\
    & \cdot & \cdot & \cdot & \cdot & \cdot & \cdot & \cdot & \textcolor{red}{\mathbf{s_y}} & \textcolor{red}{\mathbf{c_y}} & \cdot & \cdot & \cdot & \cdot & \cdot & \cdot & \cdot &\\
    & \cdot & \cdot & \cdot & \cdot & \cdot & \cdot & \cdot & \cdot & \cdot & \cdot & \cdot & \textcolor{blue}{\mathbf{-c_y}} & \textcolor{blue}{\mathbf{s_y}} & \cdot & \cdot & \cdot &\\
    & \cdot & \cdot & \cdot & \cdot & \cdot & \cdot & \cdot & \cdot & \cdot & \textcolor{blue}{\mathbf{-s_y'}} & \textcolor{blue}{\mathbf{-c_y'}} & \cdot & \cdot & \cdot & \cdot & \cdot &\\
    & \cdot & \cdot & \cdot & \cdot & \cdot & \cdot & \cdot & \cdot & \cdot & \cdot & \cdot & \cdot & \cdot & \textcolor{red}{\mathbf{c_y'}} & \textcolor{red}{\mathbf{-s_y'}} & \cdot &\\
    & \cdot & \cdot & \cdot & \cdot & \cdot & \cdot & \cdot & \cdot & \cdot & \cdot & \cdot & \textcolor{red}{\mathbf{s_y}} & \textcolor{red}{\mathbf{c_y}} & \cdot & \cdot & \cdot &\\
    & \cdot & \cdot & \cdot & \cdot & \cdot & \cdot & \cdot & \cdot & \cdot & \cdot & \cdot & \cdot & \cdot & \cdot & \cdot & -1 &\\
    & \cdot & \cdot & \cdot & \cdot & \cdot & \cdot & \cdot & \cdot & \cdot & \cdot & \cdot & \cdot & \cdot & -s_y' & -c_y' & \cdot &\\
    \end{bmatrix}.
\end{equation}
\end{widetext}

%Note that because $[S,U]=0$, $\{S,O\}=0$ implies $\{S,UOU^{\dag}\} = U\{S,O\}U^{\dag} = 0.$ Thus each term in $\mathbb{B}$ anticommutes with the symmetry $S$. Recall that our goal is to solve for operators satisfying the $0/\pi$-mode condition: $UO U^{\dag} = O$, $UOU^{\dag} = -O$ respectively. Equivalently, $F(O) = O$, $F(O) = -O$ respectively.

Given a general linear combination of elements in $\mathbb{B}$, $x = a_1\Y_{11} + b_1\Z_{11} + a_2 \X_{11}\Y_{12} + b_2 \X_{11}\Z_{12} + ... \;$, we let its column vector representation be $\mathbf{x}$:
\[
    \mathbf{x} = 
    \begin{bmatrix}
    a_1\\
    b_1\\
    a_2\\
    b_2\\
    \vdots\\
    a_{N_y}\\
    b_{N_y}\\
    \end{bmatrix}.
\]
A ZM is then found by solving the eigenvalue equation $\mathbf{F}\mathbf{x} = +\mathbf{x}$, i.e.
\begin{widetext}
\begin{equation}
    \begin{bmatrix}
    & \cdot & c_y' & -s_y' & \cdot & \cdot & \cdot & \cdot & \cdot & \cdot & \cdot & \cdot & \cdot & \cdot & \cdot & \cdot & \cdot &\\
    & 1 & \cdot & \cdot & \cdot & \cdot & \cdot & \cdot & \cdot & \cdot & \cdot & \cdot & \cdot & \cdot & \cdot & \cdot & \cdot &\\
    & \cdot & \cdot & \cdot & \textcolor{blue}{\mathbf{-c_y}} & \textcolor{blue}{\mathbf{s_y}} & \cdot & \cdot & \cdot & \cdot & \cdot & \cdot & \cdot & \cdot & \cdot & \cdot & \cdot &\\
    & \cdot & \textcolor{blue}{\mathbf{-s_y'}} & \textcolor{blue}{\mathbf{-c_y'}} & \cdot & \cdot & \cdot & \cdot & \cdot & \cdot & \cdot & \cdot & \cdot & \cdot & \cdot & \cdot & \cdot &\\
    & \cdot & \cdot & \cdot & \cdot & \cdot & \textcolor{red}{\mathbf{c_y'}} & \textcolor{red}{\mathbf{-s_y'}} & \cdot & \cdot & \cdot & \cdot & \cdot & \cdot & \cdot & \cdot & \cdot &\\
    & \cdot & \cdot & \cdot & \textcolor{red}{\mathbf{s_y}} & \textcolor{red}{\mathbf{c_y}} & \cdot & \cdot & \cdot & \cdot & \cdot & \cdot & \cdot & \cdot & \cdot & \cdot & \cdot &\\
    & \cdot & \cdot & \cdot & \cdot & \cdot & \cdot & \cdot & \textcolor{blue}{\mathbf{-c_y}} & \textcolor{blue}{\mathbf{s_y}} & \cdot & \cdot & \cdot & \cdot & \cdot & \cdot & \cdot &\\
    & \cdot & \cdot & \cdot & \cdot & \cdot & \textcolor{blue}{\mathbf{-s_y'}} & \textcolor{blue}{\mathbf{-c_y'}} & \cdot & \cdot & \cdot & \cdot & \cdot & \cdot & \cdot & \cdot & \cdot &\\
    & \cdot & \cdot & \cdot & \cdot & \cdot & \cdot & \cdot & \cdot & \cdot & \textcolor{red}{\mathbf{c_y'}} & \textcolor{red}{\mathbf{-s_y'}} & \cdot & \cdot & \cdot & \cdot & \cdot &\\
    & \cdot & \cdot & \cdot & \cdot & \cdot & \cdot & \cdot & \textcolor{red}{\mathbf{s_y}} & \textcolor{red}{\mathbf{c_y}} & \cdot & \cdot & \cdot & \cdot & \cdot & \cdot & \cdot &\\
    & \cdot & \cdot & \cdot & \cdot & \cdot & \cdot & \cdot & \cdot & \cdot & \cdot & \cdot & \textcolor{blue}{\mathbf{-c_y}} & \textcolor{blue}{\mathbf{s_y}} & \cdot & \cdot & \cdot &\\
    & \cdot & \cdot & \cdot & \cdot & \cdot & \cdot & \cdot & \cdot & \cdot & \textcolor{blue}{\mathbf{-s_y'}} & \textcolor{blue}{\mathbf{-c_y'}} & \cdot & \cdot & \cdot & \cdot & \cdot &\\
    & \cdot & \cdot & \cdot & \cdot & \cdot & \cdot & \cdot & \cdot & \cdot & \cdot & \cdot & \cdot & \cdot & \textcolor{red}{\mathbf{c_y'}} & \textcolor{red}{\mathbf{-s_y'}} & \cdot &\\
    & \cdot & \cdot & \cdot & \cdot & \cdot & \cdot & \cdot & \cdot & \cdot & \cdot & \cdot & \textcolor{red}{\mathbf{s_y}} & \textcolor{red}{\mathbf{c_y}} & \cdot & \cdot & \cdot &\\
    & \cdot & \cdot & \cdot & \cdot & \cdot & \cdot & \cdot & \cdot & \cdot & \cdot & \cdot & \cdot & \cdot & \cdot & \cdot & -1 &\\
    & \cdot & \cdot & \cdot & \cdot & \cdot & \cdot & \cdot & \cdot & \cdot & \cdot & \cdot & \cdot & \cdot & -s_y' & -c_y' & \cdot &\\
    \end{bmatrix}
    \begin{bmatrix}
    a_1\\
    b_1\\
    a_2\\
    b_2\\
    a_3\\
    b_3\\
    \vdots\\
    \vdots\\
    \vdots\\
    \vdots\\
    \vdots\\
    a_{N_y}\\
    b_{N_y}\\
    \end{bmatrix}
    = +
    \begin{bmatrix}
    a_1\\
    b_1\\
    a_2\\
    b_2\\
    a_3\\
    b_3\\
    \vdots\\
    \vdots\\
    \vdots\\
    \vdots\\
    \vdots\\
    a_{N_y}\\
    b_{N_y}\\
    \end{bmatrix}.
    \label{zmorpm}
\end{equation}
\end{widetext}
Let us evaluate this matrix multiplication one term at a time. First, the top boundary terms yield
\begin{equation}
\begin{split}
        c_y'b_1 - s_y'a_2 &= a_1\\
        a_1 &= b_1.
        \label{Topeq}
\end{split}
\end{equation}
In the bulk, we have two recurring sets of equations colored by blue and red.
When $n$ is even (blue),
\begin{equation}
\begin{split}
        -c_y b_n + s_y a_{n+1} &= a_n\\
        -s_y'b_{n-1} - c_y'a_n &= b_n,
        \label{Bulkeqeven}
\end{split}
\end{equation}
and when $n$ is odd (red),
\begin{equation}
\begin{split}
        c_y' b_n - s_y' a_{n+1} &= a_n\\
        s_y b_{n-1} + c_y a_n &= b_n.
        \label{Bulkeqodd}
\end{split}
\end{equation}
Finally, at the bottom of the lattice, 
\begin{equation}
\begin{split}
        -b_{N_y} &= a_{N_y}\\
        -s_y' a_{N_y} - c_y' b_{N_y} &= b_{N_y}.
        \label{Boteq}
\end{split}
\end{equation}

%Order of recurrence relation is a bit weird. As an illustration, let's demo with first six equations:

%\begin{equation}
%\begin{split}
%        c_y'b_1 - s_y'a_2 &= a_1\\
%        a_1 &= b_1\\
%        -c_y b_2 + s_y a_3 &= a_2\\
%        -s_y'b_1 - c_y'a_2 &= b_2\\
%        c_y' b_3 - s_y' a_4 &= a_3\\
%        s_y b_2 + c_y a_3 &= b_3.
%\end{split}
%\end{equation}
Equations~(\ref{Bulkeqeven}) and (\ref{Bulkeqodd}) can be arranged into the matrix equations
%\begin{align}
%    \begin{bmatrix}
%    s_y' & 0\\
%    c_y' & 1\\
%    \end{bmatrix}
%    \begin{bmatrix}
%    a_n\\
%    b_n\\
%    \end{bmatrix}
%    &=
%    \begin{bmatrix}
%    -1 & c_y'\\
%    0 & -s_y'\\
%    \end{bmatrix}
%    \begin{bmatrix}
%    a_{n-1}\\
%    b_{n-1}
%    \end{bmatrix}\\
%    \begin{bmatrix}
%    s_y & 0\\
%    -c_y & 1\\
%    \end{bmatrix}
%    \begin{bmatrix}
%    a_{n+1}\\
%    b_{n+1}\\
%    \end{bmatrix}
%    &=
%    \begin{bmatrix}
%    1 & c_y\\
%    0 & s_y\\
%    \end{bmatrix}
%    \begin{bmatrix}
%    a_n\\
%    b_n
%    \end{bmatrix}
%\end{align}

%At a first glance, it seems that the recursion then starts from Equations 3-6, but in fact, it starts with Equations 1,3,4,6. Equation 5 belongs to the next part of the recursion. Thus at each step of the recursion we have to analyze
%\begin{equation}
%\begin{split}
%        c_y' b_{n-1} - s_y' a_n &= a_{n-1}\\
%        -c_y b_n + s_y a_{n+1} &= a_n\\
%        -s_y'b_{n-1} - c_y'a_n &= b_n\\
%        s_y b_n + c_y' a_{n+1} &= b_{n+1}.
%\end{split}
%\end{equation}
%where $n$ is taken to be even. (Of course if $n$ is taken to be odd nothing changes, except for a reindexing of the 4 equations above).

%\textcolor{red}{From here on have to be careful with $s_y,s_y'$. When either of them is zero $F$ becomes further block-diagonal, we don't even have to do anymore analysis. Thus from here on can assume both $s_y,s_y' \neq 0$. Thus $c_y,c_y' \neq 1$.}

%Rearranging, we have
\begin{align}
    \begin{bmatrix}
    a_n\\
    b_n\\
    \end{bmatrix}
    &=
    \begin{bmatrix}
    s_y' & 0\\
    c_y' & 1\\
    \end{bmatrix}^{-1}
    \begin{bmatrix}
    -1 & c_y'\\
    0 & -s_y'\\
    \end{bmatrix}
    \begin{bmatrix}
    a_{n-1}\\
    b_{n-1}\\
    \end{bmatrix}\nonumber  \\ 
    &=
    \underbrace{
    \frac{1}{s_y'}
    \begin{bmatrix}
    -1 & c_y'\\
    c_y' & -1\\
    \end{bmatrix}
    }_{M_{eo}}
    \begin{bmatrix}
    a_{n-1}\\
    b_{n-1}\\
    \end{bmatrix} \label{oddtoeven}
\end{align}
and
\begin{align}
    \begin{bmatrix}
    a_{n+1}\\
    b_{n+1}\\
    \end{bmatrix}
    &=
    \begin{bmatrix}
    s_y & 0\\
    -c_y & 1\\
    \end{bmatrix}^{-1}
    \begin{bmatrix}
    1 & c_y\\
    0 & s_y\\
    \end{bmatrix}
    \begin{bmatrix}
    a_n\\
    b_n\\
    \end{bmatrix} \nonumber \\ 
    &=
    \underbrace{
    \frac{1}{s_y}
    \begin{bmatrix}
    1 & c_y\\
    c_y & 1\\
    \end{bmatrix}
    }_{M_{oe}}
    \begin{bmatrix}
    a_n\\
    b_n\\
    \end{bmatrix}, \label{eventoodd}
\end{align}
where $n$ is an even number. The matrices $M_{eo}$ and $M_{oe}$ are the transition matrices from the odd-to-even and even-to-odd sites respectively. Combining the two equations above, we have 
%Now we have two separate strands to analyse. We shall consider just the sublattice of odd sites, and shall find that as $n \rightarrow \infty$, $a_n,b_n \rightarrow 0$. Similarly on the sublattice of even sites, as $n \rightarrow \infty$, $a_n,b_n \rightarrow 0$.
\begin{equation}
    \begin{bmatrix}
    a_n\\
    b_n\\
    \end{bmatrix}
    = 
    \underbrace{
    \frac{1}{s_ys_y'}
    \begin{bmatrix}
    c_yc_y'-1 & c_y'-c_y\\
    c_y'-c_y & c_yc_y'-1\\
    \end{bmatrix}
    }_{M_{o}}
    \begin{bmatrix}
    a_{n-2}\\
    b_{n-2}\\
    \end{bmatrix},
    \label{recrel}
\end{equation}
where $M_o \coloneqq M_{oe}M_{eo}$ and $n>1$ is odd. For convenience, we let $x_n$ denote $[a_n, b_n]^T$, i.e., $x_n = M_o x_{n-2} = M^{\frac{n-1}{2}} x_1$.

Since eigenvectors are unique up to a scaling parameter, we can fix $a_1=1$. The second line of Eq.~(\ref{Topeq}) then yields $b_1=1$ too. Observe that $x_1 = [a_1,b_1]^T = [1,1]^T$ is an eigenvector of $M$ with eigenvalue
\[
    \lambda = \frac{1}{s_ys_y'} (c_y+1)(c_y'-1).
\]
Thus 
\begin{align*}
    ||{x_n}||^2 = x_n^Tx_n &= x_1^T(M^{\frac{n-1}{2}})^T M^{\frac{n-1}{2}} x_1\\
    &= x_1^T \lambda^{\frac{n-1}{2}} \lambda^{\frac{n-1}{2}} x_1\\
    &= (\lambda^2)^{\frac{n-1}{2}} ||x_1||^2 \;\;\xrightarrow{n \rightarrow \infty} 0\\ 
    &\quad \text{if and only if} \;\;\lambda^2 <1.
\end{align*}
That is, since a physical corner ZM is obtained if $a_n$ and $b_n$ converge exponentially to zero away from $n=1$, the following condition must hold:
\[
    \lambda^2 = 
    \frac{1}{(s_ys_y')^2} (c_y+1)^2(c_y'-1)^2 = \frac{(1+c_y)(1-c_y')}{(1-c_y)(1+c_y')}<1.
\]
This is equivalent to the condition $c_y<c_y'$, i.e. $\cos(2J_y) < \cos(2J_y')$, as claimed in the main text.

At this point, the analysis above is not entirely complete since Eq.~(\ref{recrel}) only relates $a_n$ and $b_n$ with odd $n$. Nevertheless, a similar recurrence matrix equation involving $a_n$ and $b_n$ with even $n$ can be easily constructed. In this case, the base point $x_2$ can be obtained by applying Eq.~(\ref{oddtoeven}). A similar analysis as above then yields the same condition for the convergence of a ZM. Thus, the analysis for corner ZMs is complete.

A similar procedure can be applied to find the condition for the existence of PMs. We now solve the eigenvalue equation for $\mathbf{F}\mathbf{x} = -\mathbf{x}$ instead. A similar recurrence matrix equation of the form  $x_n = M_o x_{n-2}$ is obtained, where we now have
\[
    M_o =
    \frac{1}{s_ys_y'}
    \begin{bmatrix}
    c_yc_y'-1 & c_y-c_y'\\
    c_y-c_y' & c_yc_y'-1\\
    \end{bmatrix}, \quad x_1 =
    \begin{bmatrix}
    1\\
    -1\\
    \end{bmatrix}.
\]
By noting that the base point $x_1$ is again an eigenvector of $M_o$, the same analysis as above leads to the conclusion that $\cos(2J_y)<\cos(2J_y')$ is again both a necessary and sufficient condition for corner PMs to exist.

\textcolor{black}{
To show that we only have corner modes, i.e. there are actually no modes along the edges of the lattice, we again employ the same strategy. Without loss of generality, let us begin with operators restricted to the lattice site $(1,2)$ on the edge. As before, $(h,J_x)=(\pi/4,\pi/2)$, and here we further set $J_y'=0$. Evaluating $F$:
\begin{align*}
    F(\Y_{12}) &= \Z_{12}\\
    F(\Z_{12}) &= c_y\Y_{12}-s_y\X_{12}\Z_{13}\\
    F(\X_{12}\Y_{13}) &= -c_y\X_{12}\Z_{13}-s_y\Y_{12}\\
    F(\X_{12}\Z_{13}) &= -\X_{12}\Y_{13}.
\end{align*}
Thus with respect to the basis $\{\Y_{12},\Z_{12},\X_{12}\Y_{13},\X_{12}\Z_{13}\}$, the corresponding matrix analogous to that in \ref{Fmatrix} is simply
\begin{align*}
    \mathbf{F} = 
    \begin{bmatrix}
        0 & c_y & -s_y & 0\\
        1 & 0 & 0 & 0\\
        0 & 0 & 0 & -1\\
        0 & -s_y & -c_y & 0
    \end{bmatrix}
\end{align*}
(note that there is no endless propagation down the lattice). Solving for the characteristic equation $\text{det}(\lambda I - F) = 0$ gives $\lambda^4 - 2c_y\lambda^2 +1 = 0$. Thus we see that $\lambda=\pm1$ are solutions if and only if $c_y=1$, i.e. $J_y=0/\pi$. That is, a 0 ($\pi$) mode, which corresponds to $\lambda=+1$ ($\lambda=-1$), only exists at site (1,2) in the case where all the horizontal chains are effectively decoupled (at $J_y=\pi$, the vertical coupling only introduces a phase factor of $-1$ to all horizontal chains in the bulk), and the system indeed reduces to $N_y$ independent 1D chains. However, in the presence of even the slightest nontrivial coupling $J_y\neq 0,\pi$, such a 0 ($\pi$) mode is no longer present. The same is true if $J_y'\neq 0$ instead. This is consistent with the intuition elucidated in the main text that the coupling between two neighboring 1D chains in the bulk causes these edge 0 and $\pi$ modes to hybridize and disappear, leaving only those at the corners that are truly topological. 
}

%\textcolor{red}{Our numerics show that the model viewed as stack of 1D floquet SSH model, where each (horizontal) chain supports 0 and $\pi$ modes at the ends, vertical $J_y$ couplings will hybridize pairs of 0 and $\pi$ modes that are not at the corners, so no solely $0/\pi$.
%}

%This explains why there are no separate phases containing solely $0$-modes or $\pi$-modes. There is one last condition: the boundary conditions at the other end. As we go to the thermodynamic limit this gets more and more precise, since everything goes to 0, so all good. This completes our derivation of the main result.

\section{Evaluation of $\nu_0$ and $\nu_\pi$}
\label{app2}

First we define the unitary operator
\begin{equation}
    R = e^{-i \frac{\pi}{3\sqrt{3}} (\X + \Y + \Z)\otimes I} e^{-i \frac{\pi}{4} \X \Z }.
\end{equation}
The transformed BdG Hamiltonians $\tilde{H}_{1,\textbf{BdG}}(k) = R H_{1,\textbf{BdG}}(k) R^\dagger$ and $\tilde{H}_{2,\textbf{BdG}}(k) = R H_{2,\textbf{BdG}}(k) R^\dagger$ respect a chiral symmetry under $ \mathcal{C}=\Z \otimes I$. In this case, the $F(k)$ matrix defined in the main text can be written in the form
\begin{equation}
    F(k)=e^{-i\frac{H_{2,\textbf{BdG}}(k)}{2}}e^{-i\frac{H_{1,\textbf{BdG}}(k)}{2}}=\left[\begin{array}{cc}
        A(k) & B(k)  \\
        C(k) & D(k)
    \end{array} \right] \;,
\end{equation}
where $A(k)$, $B(k)$, $C(k)$, and $D(k)$ are $2\times 2$ matrices. In particular,
%\begin{equation}
%    \begin{split}
%        e^{-iH_{1,\textbf{BdG}}(k)} = &\frac{1}{\sqrt{2}} - \frac{i}{\sqrt{2}} %\Y \otimes \Z \\
%        e^{-iH_{2,\textbf{BdG}}(k)} = 
%        &\frac{1}{2}(-s_y'+c_y)I \otimes I + \frac{1}{2}(-ic_y'+is_yc_k)\Y \otimes \Y +\\
%        &\frac{1}{2}(s_y'+c_y)\Z \otimes \Z + \frac{1}{2}(-ic_y'-is_yc_k)\X \otimes \X +\\ 
%        &\frac{1}{2}(-is_ys_k)\Y \otimes \X + \frac{1}{2}(-is_ys_k)\X \otimes \Y.
%    \end{split}
%\end{equation}
%As per Asboth's prescription, multiply both terms to get
%\begin{equation}
%    \begin{split}
%        2\sqrt{2} \cdot F(k) = 
%        2\sqrt{2} \cdot e^{-iH_{2,\textbf{BdG}}(k)}e^{-iH_{1,\textbf{BdG}}(k)} = 
%        &\;(-ic_y'+is_yc_k)\Y \otimes \Y + (-ic_y'-is_yc_k)\X \otimes \X +\\
%        &\;(-is_ys_k)\Y \otimes \X + (-is_ys_k)\X \otimes \Y +\\ 
%        &\;(is_y'-ic_y)\Y \otimes \Z + (-s_y'-c_y)\X \otimes I \\
%        &\quad+\\
%        &\;(-s_y'+c_y)I \otimes I + (s_y'+c_y)\Z \otimes \Z +\\
%        &\;(-ic_y'+is_yc_k)I \otimes \X + (-c_y'-s_yc_k)\Z \otimes \Y +\\ 
%        &\;(is_ys_k)I \otimes \Y + (s_ys_k)\Z \otimes \X
%    \end{split}
%\end{equation}
\begin{equation}
\begin{split}
    B(k) &= \frac{1}{\sqrt{2}}
    \begin{bmatrix}
    -c_y & -s_ys_k-is_yc_k\\
    -ic_y' & -s_y'
    \end{bmatrix} \\
    D(k) &= \frac{1}{\sqrt{2}}
    \begin{bmatrix}
    -s_y' & -ic_y'\\
    -s_ys_k+is_yc_k & c_y
    \end{bmatrix},
\end{split}
\end{equation}
where now we use the shorthand notation $c_y=\cos(J_y)$, $s_y=\sin(J_y)$, $c_y'=\cos(J_y')$, and $s_y'=\sin(J_y')$. Note that this is \emph{different} from the shorthand notation used in Appendix~\ref{app1}.
%\[
%    B(k) = \frac{1}{\sqrt{2}}
%    \begin{bmatrix}
%    -c_y & -s_ys_k-is_yc_k\\
%    -ic_y' & -s_y'
%    \end{bmatrix},
%    \;\;\text{so}\;\; \text{det}\; B(k) =\; \frac{1}{2}\left( c_ys_y' + s_yc_y'e^{-ik} \right) .
%\]
%From the next three lines, we get 
%\[
%    D(k) = \frac{1}{\sqrt{2}}
%    \begin{bmatrix}
%    -s_y' & -ic_y'\\
%    -s_ys_k+is_yc_k & c_y
%    \end{bmatrix},
%    \;\;\text{so}\;\; \text{det}\; D(k) =\; \frac{1}{2}\left( -c_ys_y' - %s_yc_y'e^{ik} \right) .
%\]

According to Ref. \cite{Asboth_etal}, ZMs and PMs can respectively be evaluated from the winding numbers $\nu[B] = \frac{1}{2\pi i} \int_{-\pi}^{\pi} dk \frac{d}{dk} \text{ln} \;[\text{det}\;B(k)]$ and $\nu[D] = \frac{1}{2\pi i} \int_{-\pi}^{\pi} dk \frac{d}{dk} \text{ln} \;[\text{det}\;D(k)]$. We now evaluate these integrals analytically. 

Consider $\text{det}\; B(k)$. This function of $k$ can be viewed as a composition of functions $f$ and $z$, where $f: \mathbb{C} \longrightarrow \mathbb{C}$ is a complex function and $z: \mathbb{R} \longrightarrow \mathbb{C}$ traces a path in the complex plane parametrized by $k$. That is, $\text{det}\; B(k) = (f\circ z)(k)$. Here, $f(z) = z_{0,B} + z$, where $z_{0,B}=c_ys_y'$ and $z(k)=s_yc_y'e^{-ik}$. We have ignored the $\frac{1}{2}$-factor, which does not affect the integral. So,
\begin{equation}
    \begin{split}
        \nu[B] &= \frac{1}{2\pi i} \int_{-\pi}^{\pi} dk \frac{d}{dk} \text{ln} \;[\text{det}\;B(k)]\\
        &= \frac{1}{2\pi i} \int_{-\pi}^{\pi} dk \frac{d}{dk} \text{ln} \;f(z(k))\\
        &= \frac{1}{2\pi i} \int_{-\pi}^{\pi} \frac{f'(z(k))}{f(z(k))} z'(k) dk\\
        &= \frac{1}{2\pi i} \oint_C \frac{f'(z)}{f(z)} dz,
    \end{split}
\end{equation}
where $C$ is the (negatively oriented) contour centered at the origin and with radius $|s_yc_y'|$. Next we apply the Argument Principle from Complex Analysis, which asserts that so long $f$ is meromorphic within $C$, the value of the integral above is given simply by $Z-P$, where $Z,P$ are respectively the number of zeros and poles of $f$ within $C$.  

\begin{center}
\begin{figure}
    \includegraphics[scale=0.5]{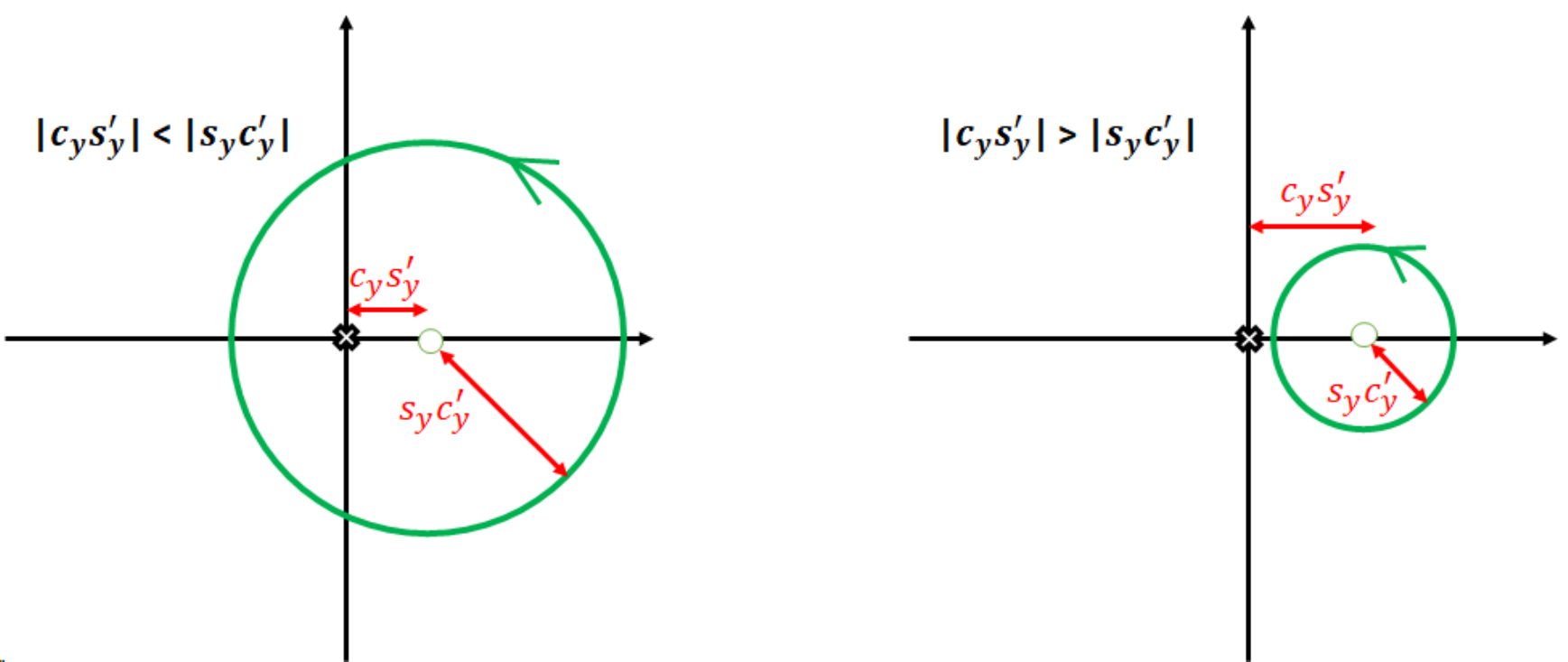}
    \caption{The winding number $\nu[B]$ is only nonzero if the integration contour encloses the origin.}
    \label{fig: contour}
\end{figure}    
\end{center}

Now clearly $f$ is analytic everywhere, so there are no poles. On the other hand, there is a single zero if $z_{0,B} \in G_0\equiv \{z\;|\; |z|<|s_yc_y'|\}$, i.e., the integration contour encloses the origin (see Fig.~\ref{fig: contour}), and no zero otherwise. Therefore, we find  

%How many zeros of $f$ are there within the interior of $C$, that is, within the circular domain $D = \{z\;|\; |z|<|s_yc_y'|\}$? That is equivalent to asking when is $z_{0,B} \in D$. When $z_{0,B} \in D$, $f$ has a zero in $D$; when $z_{0,B} \notin D$, $f$ has no zeros in $D$. We infer:
\begin{equation}
    \begin{split}
        \text{ZMs exist} \Longleftrightarrow \nu[B]=1 &\Longleftrightarrow z_{0,B} \in G_0\\ &\Longleftrightarrow |c_ys_y'|<|s_yc_y'|\\
        &\Longleftrightarrow |\text{tan}\;J_y| > |\text{tan}\;J_y'|\\
        &\Longleftrightarrow \text{cos}\;2J_y < \text{cos}\;2J_y',
    \end{split}
\end{equation}
where the last equivalence was obtained using the trigonometric identity $\text{tan}^2\;\theta + 1 = \text{sec}^2\;\theta$. 

To determine the presence of PMs, we repeat the same analysis above to find $\nu[D]$. In particular, noting that \text{det}\;$D(k)=z_{0,D}+z(k)$, where $z_{0,D} = -c_ys_y'$ and $z(k) = -s_yc_y'e^{ik}$, we obtain 
\begin{equation}
    \begin{split}
        \text{PMs exist} \Longleftrightarrow \nu[D]=1 &\Longleftrightarrow z_{0,D} \in G_\pi ( = G_0) \\ &\Longleftrightarrow |c_ys_y'|<|s_yc_y'|\\
        &\Longleftrightarrow |\text{tan}\;J_y| > |\text{tan}\;J_y'|\\
        &\Longleftrightarrow \text{cos}\;2J_y < \text{cos}\;2J_y'.
    \end{split}
\end{equation}

%\clearpage
%\newpage
%\end{widetext}

%\bibliographystyle{unsrt}
%\bibliography{main}

\end{document}